\documentclass[a4paper,11pt]{article}
\pdfoutput=1
\usepackage{jcappub,xcolor,graphicx,epsf,bm,amsmath,amsfonts,amssymb,epstopdf,natbib,hyperref,color,verbatim,multirow,bm}
\usepackage[T1]{fontenc}
\hypersetup{colorlinks=true,urlcolor=blue,citecolor=blue,linkcolor=blue,menucolor=blue,anchorcolor=blue,filecolor=blue}
\DeclareMathOperator{\arcsinh}{arcsinh}

\begin{document}
\title{Testing scale-invariant inflation against cosmological data}

\author[a,b,c]{Chiara Cecchini}
\author[d]{Mariaveronica De Angelis}
\author[d]{William Giar\`{e}}
\author[a,b]{Massimiliano Rinaldi}
\author[a,b]{and Sunny Vagnozzi}
\affiliation[a]{Department of Physics, University of Trento, Via Sommarive 14, 38122 Povo (TN), Italy}
\affiliation[b]{Trento Institute for Fundamental Physics and Applications (TIFPA)-INFN, Via Sommarive 14, 38122 Povo (TN), Italy}
\affiliation[c]{Department of Theoretical Physics, CERN, Esplanade des Particules 1, 1211 Geneva 23, Switzerland}
\affiliation[d]{School of Mathematics and Statistics, University of Sheffield, Hounsfield Road, Sheffield S3 7RH, United Kingdom}
\emailAdd{chiara.cecchini@unitn.it}
\emailAdd{mdeangelis1@sheffield.ac.uk}
\emailAdd{w.giare@sheffield.ac.uk}
\emailAdd{massimiliano.rinaldi@unitn.it}
\emailAdd{sunny.vagnozzi@unitn.it}

\abstract{There is solid theoretical and observational motivation behind the idea of scale-invariance as a fundamental symmetry of Nature. We consider a recently proposed classically scale-invariant inflationary model, quadratic in curvature and featuring a scalar field non-minimally coupled to gravity. We go beyond earlier analytical studies, which showed that the model predicts inflationary observables in qualitative agreement with data, by solving the full two-field dynamics of the system -- this allows us to corroborate previous analytical findings and set robust constraints on the model's parameters using the latest Cosmic Microwave Background (CMB) data from \textit{Planck} and \textit{BICEP/Keck}. We demonstrate that scale-invariance constrains the two-field trajectory such that the effective dynamics are that of a single field, resulting in vanishing entropy perturbations and protecting the model from destabilization effects. We derive tight upper limits on the non-minimal coupling strength, excluding conformal coupling at high significance. By explicitly sampling over them, we demonstrate an overall insensitivity to initial conditions. We argue that the model \textit{predicts} a minimal level of primordial tensor modes set by $r \gtrsim 0.003$, well within the reach of next-generation CMB experiments. These will therefore provide a litmus test of scale-invariant inflation, and we comment on the possibility of distinguishing the model from Starobinsky and $\alpha$-attractor inflation. Overall, we argue that scale-invariant inflation is in excellent health, and possesses features which make it an interesting benchmark for tests of inflation from future CMB data.}
\maketitle

\section{Introduction}
\label{sec:introduction}

Cosmic inflation, a postulated stage of quasi-de Sitter expansion in the extremely early Universe~\cite{Kazanas:1980tx,Starobinsky:1980te,Sato:1981ds,Guth:1980zm,Mukhanov:1981xt,Linde:1981mu,Albrecht:1982wi}, can now basically be considered an integral part of the standard cosmological model, despite the lack of direct, empirical evidence for the inflationary stage. The inflationary paradigm remains in very good health (despite a few potential foundational problems~\cite{Ijjas:2013vea,Ijjas:2014nta,Obied:2018sgi,Agrawal:2018own,Achucarro:2018vey,Garg:2018reu,Kehagias:2018uem,Kinney:2018nny,Ooguri:2018wrx,Palti:2019pca,Bedroya:2019tba,Geng:2019phi,Odintsov:2020zkl,Trivedi:2020wxf,Vagnozzi:2022qmc}), and is in excellent agreement with a variety of cosmological probes~\cite{Benetti:2013wla,Choudhury:2013woa,Benetti:2013cja,Martin:2013nzq,Creminelli:2014oaa,Dai:2014jja,Escudero:2015wba,Benetti:2016ycg,Gerbino:2016sgw,Guo:2017qjt,Ni:2017jxw,SantosdaCosta:2017ctv,Guo:2018uic,Ballardini:2018noo,DiValentino:2018wum,Benetti:2019kgw,Haro:2019peq,Guo:2019dui,Li:2019ipk,Braglia:2020fms,Cicoli:2020bao,SantosdaCosta:2020dyl,Odintsov:2020mkz,Vagnozzi:2020gtf,Vagnozzi:2020rcz,Vagnozzi:2020dfn,Stein:2021uge,Forconi:2021que,dosSantos:2021vis,Benetti:2021uea,Ballardini:2022wzu,Ye:2022afu,Antony:2022ert,Chen:2022dyq,Shiravand:2022ccb,Gangopadhyay:2022vgh,RoyChoudhury:2022rva,Montefalcone:2022owy,Stein:2022cpk,Cabass:2022oap,Montefalcone:2022jfw,NANOGrav:2023hvm,Vagnozzi:2023lwo,Niu:2023bsr,Unal:2023srk,Choudhury:2023kam,Firouzjahi:2023lzg,HosseiniMansoori:2023mqh,Ben-Dayan:2023lwd,Jiang:2023gfe,Frosina:2023nxu,Choudhury:2023hfm,Bhattacharya:2023ysp,Oikonomou:2023bli,Bostan:2023ped,Choudhury:2023fwk,Choudhury:2023fjs,Giare:2024sdl,Choudhury:2024one}, while a substantial improvement in the determination of inflationary parameters, and the detection of the ``smoking gun'' signature of inflationary gravitational waves (GWs), is among the key goals of various upcoming surveys~\cite{Matsumura:2013aja,CMB-S4:2016ple,SimonsObservatory:2018koc,SimonsObservatory:2019qwx,Giovannini:2019oii,Giare:2020vhn,Campeti:2020xwn,CMB-S4:2020lpa,LiteBIRD:2022cnt,LiteBIRD:2023zmo}. While on the observational side the health of the inflationary paradigm endures~\cite{Chowdhury:2019otk}, on the theory side the situation is arguably less simple. Precision measurements of the Cosmic Microwave Background (CMB) have in fact long ruled out some of the simplest (minimally coupled) single-field monomial models, whose predictions for the amplitude of inflationary tensor modes exceed by far current constraints on the tensor-to-scalar ratio $r$, once the tilt of the scalar power spectrum $n_s$ is fixed within the observationally allowed range~\cite{Planck:2018jri}. This impasse has triggered various new lines of investigation partially shifting the focus from a particle-like origin for inflation to quantum effects in the gravitational sector and in vacuum: one intriguing line of research in this direction embraces the idea of (global) scale-invariance as a fundamental symmetry of Nature.

The idea of scale-invariance (or scale symmetry) has a long and rich history, and has been discussed in a variety of theoretical contexts. At the classical level, scale symmetry demands the absence of explicit, fundamental scales in the action: this restricts the only allowed operators in $3+1$ dimensions to be dimension-$4$ ones (we will be more explicit later), significantly increasing the predictivity of the theory. Nevertheless, explicit mass scales typically emerge due to the running of couplings. At the quantum level (see e.g.\ Refs.~\cite{Wetterich:1987fm,Shaposhnikov:2008xi,Nakayama:2010zz,Mooij:2018hew,Wetterich:2019qzx}), scale symmetry requires the quantum effective action to not contain any intrinsic dimensionful parameter. Quantum scale symmetry has been proposed as a new theoretical guiding principle for UV-complete theories going beyond renormalizability~\cite{Wetterich:2020cxq}. While theories displaying fundamental scale-invariance are naturally renormalizable, they further admit a scale-free formulation of the effective action (i.e.\ the dependence on the renormalization scale $k$ can always be absorbed into a definition of scaling fields). As a consequence, all the relevant parameters of the theory (encoding deviations from the exact scaling solution) vanish, endowing the theory itself with an extremely high predictive power~\cite{Wetterich:2020cxq}.

If scale-invariance as a fundamental (either classical or quantum) symmetry of Nature enjoys strong theoretical motivation, equally strong motivation for the idea exists from the phenomenological and observational perspectives, both in the fields of particle physics and cosmology. On the particle side, an interesting motivation for scale-invariance is related to the naturalness problem which, at its heart, is connected to divergent quantum corrections to parameters associated to super-renormalizable (mass dimension $<4$) Lagrangian terms, such as the vacuum energy and the Higgs mass~\cite{Craig:2022eqo}. This might be seen as an indication that, barring other protection mechanisms, Nature may (for want of a better term) prefer dimension-$4$ operators: in this case, no explicit mass scale appears at the classical level. On the cosmological side the inferred spectral index $n_s$, combined with the observed small amount of anisotropies and stringent limits on inflationary tensor modes~\cite{Planck:2018jri}, appears to require models with unusually flat potentials, often even for super-Planckian field values. This raises the question of how the flatness of the potential is preserved against radiative corrections (once more barring protection mechanisms). These features can more easily be accommodated if only dimension-$4$ terms, thereby controlled by dimensionless couplings, are allowed by Nature. The above phenomenological considerations, coupled with the earlier theoretical ones, lend strong support to the idea of scale-invariance being a good candidate symmetry for a fundamental theory of Nature, which should not contain explicit mass or length scales: with no claims as to completeness, ideas along this direction have been explored in various works, see e.g.\ Refs.~\cite{Coleman:1973jx,Englert:1976ep,Cooper:1981byv,Hempfling:1996ht,Meissner:2006zh,Foot:2007as,Foot:2007iy,Finelli:2007wb,Foot:2010av,Tronconi:2010pq,Englert:2013gz,Heikinheimo:2013fta,Hambye:2013dgv,Khoze:2013uia,Holthausen:2013ota,Einhorn:2014gfa,Salvio:2015cja,Edery:2015wha,Rinaldi:2015yoa,Einhorn:2015lzy,Einhorn:2016mws,Ghoshal:2017egr,Cosme:2018nly,Burrage:2018dvt,Salvio:2019wcp,Gakis:2019rdd,Ghoshal:2020lfd,Salvio:2020axm,Ferreira:2021ctx,Barman:2021lot,Ghoshal:2022hyc,Barman:2022njh,Alvarez-Luna:2022hka,Salvio:2023qgb,Giani:2023tai}.

Returning to inflation, the fact that scale-invariance may play an important role can be better appreciated by inspecting the reasons behind the success of the Starobinsky model~\cite{Starobinsky:1980te}, currently among the models which best fit the data, and whose predictions in the $n_s$-$r$ plane are considered by many to be a key experimental target. Starobinsky inflation relies on a modification of the vacuum Einstein-Hilbert action, obtained by adding a term proportional to $R^2$, i.e.\ quadratic in the Ricci scalar. The model can be recast as a scalar-tensor theory, through a conformal transformation of the metric from the Jordan Frame (JF) to the Einstein Frame (EF): this results in a scalar potential which is extremely flat at large field values, in turn explaining the excellent agreement with cosmological observations.~\footnote{Similar considerations can be extended to Starobinsky-like inflationary models, which rely on a scalar field $\phi$ non-minimally coupled to the Ricci scalar, via a coupling of the form $-1/2f(\phi)R$~\cite{Fakir:1990eg,Kaiser:1994vs,Burgess:2009ea,Barbon:2009ya,Burgess:2014lza,Giudice:2014toa,Joergensen:2014rya,Myrzakulov:2015qaa,vandeBruck:2015xpa,Salvio:2015kka,Bahamonde:2015shn,Myrzakul:2015gya,Kaneda:2015jma,Calmet:2016fsr,Salvio:2016vxi,Sebastiani:2016ras,Wang:2017fuy,Ema:2017rqn,Mathew:2017lvh,Hammad:2017svs,Cheraghchi:2018sgq,He:2018gyf,Gorbunov:2018llf,Kleidis:2018cdx,Enckell:2018hmo,Canko:2019mud,Bezrukov:2019ylq,Ema:2019fdd,Das:2020kff,Zhang:2021ppy,Panda:2022can,Daniel:2022ppp,Dioguardi:2022oqu,AlHallak:2023wsx,Odintsov:2023weg}. Particular choices of this non-minimal coupling can be understood as an unavoidable requirement for the renormalization of the stress-energy tensor of Standard Model fields when quantized in a gravitational background. The best known model in this class is of course Higgs inflation~\cite{Bezrukov:2007ep,Bezrukov:2010jz,Rubio:2018ogq}.} The origin of such a flatness is rooted in the $R^2$ term appearing in the JF action, and dominating the dynamics during inflation by virtue of its large (dimensionless) coefficient. The Starobinsky model is a particular case of $f(R)$ model motivated by quantum corrections to gravity~\cite{Buchdahl:1970ldb,Sotiriou:2008rp,DeFelice:2010aj}. One can attempt to define the functional form of $f(R)$ that best accommodates inflationary observables consistent with data, via reconstruction techniques based on the measured spectral indices: the results of such an exercise clearly indicate a preference for a quadratic term in the Ricci scalar~\cite{Rinaldi:2014gua} (see also Refs.~\cite{Oikonomou:2014jua,Myrzakulov:2015fra,Sebastiani:2015kfa,Odintsov:2017fnc,Odintsov:2018ggm,Chiba:2018cmn,Nojiri:2019dqc,Odintsov:2020fxb,Odintsov:2021wjz,Giacomozzi:2024eyo}). The fundamental reason why the Starobinsky model and models close thereto perform extremely well is, in essence, scale-invariance. Being $R^2$ a dimension-4 term, its coefficient is dimensionless and the quadratic part of the action is therefore invariant under Weyl rescaling $\bar{g}_{\mu\nu}(x)=g_{\mu\nu}(\ell x)$ for positive constants $\ell$. As a consequence, when the curvature is large the quadratic term dominates and drives inflation away from an unstable de Sitter fixed point. At sufficiently late times the standard Einstein-Hilbert term linear in $R$ overcomes the quadratic term, (approximate) scale-invariance is broken, and inflation ends.

Driven by the above considerations, it appears reasonable to push these ideas further and entertain the possibility that before/at the start of inflation, the Universe starts out in a scale-invariant state, with the subsequent evolution breaking this symmetry while driving inflation. These are the key ideas presented by one of us in Ref.~\cite{Rinaldi:2015uvu}, which considered an action containing all relevant scale-invariant terms obtained by combining the Ricci scalar and a real scalar field. On a flat Friedmann-Lema\^{i}tre-Robertson-Walker (FLRW) background metric, the equations of motion show that the solution interpolates between two de Sitter spaces, one being an attractor and the other a saddle point. During the transition between the two regimes, the scalar field stabilizes around a non-zero value which can be associated to a mass scale, and ultimately identified with the Planck mass. In this picture inflation is then accompanied by a spontaneous/dynamical breaking of scale-invariance and the emergence of a mass scale not present in the original action. Subsequent works have established analytically that the spectral indices predicted by this model fall well within current experimental constraints~\cite{Tambalo:2016eqr,Ghoshal:2022qxk,Rinaldi:2023mdf}, even when one-loop quantum corrections are considered~\cite{Vicentini:2019etr}. Scale-invariance and inflation have also been studied in other papers. For instance, Refs.~\cite{Garcia-Bellido:2011kqb,Ferreira:2018qss} point out how the dynamics in a two-field scale-invariant theory in the EF is basically driven by only one field, precisely as a consequence of scale-invariance. In turn, this guarantees the (desirable) absence of isocurvature perturbations. For other works on the role of scale-invariance in inflation, we refer the reader for instance to Refs.~\cite{Bars:2013yba,Salvio:2014soa,Kannike:2014mia,Csaki:2014bua,Rinaldi:2014gha,Kannike:2015apa,Bamba:2015uxa,Kannike:2015kda,Farzinnia:2015fka,Ferreira:2016vsc,Karananas:2016kyt,Ferreira:2016wem,Salvio:2017xul,Casas:2017wjh,Ferreira:2018itt,Benisty:2018fja,Kubo:2018kho,Casas:2018fum,Shaposhnikov:2018nnm,Gialamas:2020snr,vandeBruck:2021xkm,Gialamas:2021enw,Rosenlyst:2022jxj,Cecchini:2023bqu,Karananas:2023zgg,Aoki:2024jhr}.~\footnote{Complementary aspects on black holes and other compact objects in scale-invariant theories of gravity were recently studied in Refs.~\cite{Cognola:2015wqa,Cognola:2015uva,Bonanno:2019rsq,Ferreira:2019ywk,Aydemir:2020xfd,Dioguardi:2020nxr,Bonanno:2021zoy,Silveravalle:2022lid,Silveravalle:2022wij,Boudet:2022fke}.}

The earlier works of Refs.~\cite{Tambalo:2016eqr,Vicentini:2019etr,Ghoshal:2022qxk,Rinaldi:2023mdf} have analytically confirmed that inflation in a scale-invariant Universe as discussed in Ref.~\cite{Rinaldi:2015uvu} is a viable model for the origin of structure as far as inflationary observables are concerned. At the same time, much work remains to be done on various fronts, including but not limited to:
\begin{enumerate}
\item corroborating the earlier analytical results by numerically solving the complete field equations throughout the entire period of inflation -- this can be challenging because, at least on paper (as is clear once one moves to the EF), this is a two-field model;
\item determining the exact constraints set by current cosmological data on the model's parameters, since the comparison to observational data performed in earlier works is somewhat qualitative, aimed at identifying viable benchmark points in parameter space;
\item an analysis along the above lines should ideally also sample on the initial conditions for the dynamical degrees of freedom, in order to investigate whether inflation can occur for generic (or, conversely, fine-tuned) values thereof;
\item similarly, rather than considering priors on observables (such as $n_s$ and $r$), it would be interesting to consider priors on the fundamental model parameters;
\item assessing the predicted level of isocurvature perturbations, eventually corroborating the results of Refs.~\cite{Garcia-Bellido:2011kqb,Ferreira:2018qss};
\item computing the predicted level of non-Gaussianity;
\item finally, exploring the extent to which the model can be distinguished from competing models, in particular Starobinsky inflation, is in order.
\end{enumerate}
The goal of the present work is precisely that of addressing the above points, in order to further cement the observational viability of scale-invariant inflation. Most of the above points, but especially 1--4, are tackled thanks to the numerical method recently developed by two of us in Ref.~\cite{Giare:2023kiv}, specifically designed to study generic multi-field inflationary models whose field space metric is potentially non-trivial. Our code solves the full numerical dynamics of the model and calculates precise predictions for various observable quantities beyond $n_s$ and $r$, while allowing us to efficiently explore the impact of initial conditions. Moreover, it is interfaced with Boltzmann solvers and a Monte Carlo sampler which allows us to easily compare the model's predictions against current CMB observations (for which an efficient analytical -- compressed -- likelihood is used), thereby setting precise constraints on the model's parameters as per points 2.\ and 3.\ above. Ultimately, our results corroborate the validity of scale-invariant inflation, and pave the way for further tests of the potentially important role of scale-invariance in cosmology.

The rest of this paper is then organized as follows. In Sec.~\ref{sec:model} we review the basic features of the scale-invariant model of inflation we consider. Various analytical aspects of the model, most of which will be useful for the later analysis, are presented in Sec.~\ref{sec:theory}: in particular, in Sec.~\ref{subsec:entropy} we introduce the methodology and prove that entropy perturbations vanish due to scale symmetry, whereas in Sec.~\ref{subsec:redefinition} we compute all the relevant quantities useful for the later numerical implementation, and in Sec.~\ref{subsec:nonGaussianity} we compute the level of local non-Gaussianity predicted by the model. The results of our numerical analysis are discussed in Sec.~\ref{sec:cosmoconstraints}, with Sec.~\ref{subsec:methods} devoted to reviewing the numerical method, Sec.~\ref{subsec:results} to discussing the constraints we obtain, and Sec.~\ref{subsec:discussion} to comparing our observational predictions against those of other benchmark models. We close in Sec.~\ref{sec:conclusions} by drawing concluding remarks.

\section{Scale-invariant inflation}
\label{sec:model}

\begin{figure}
\centering 
\includegraphics[width=10.5cm]{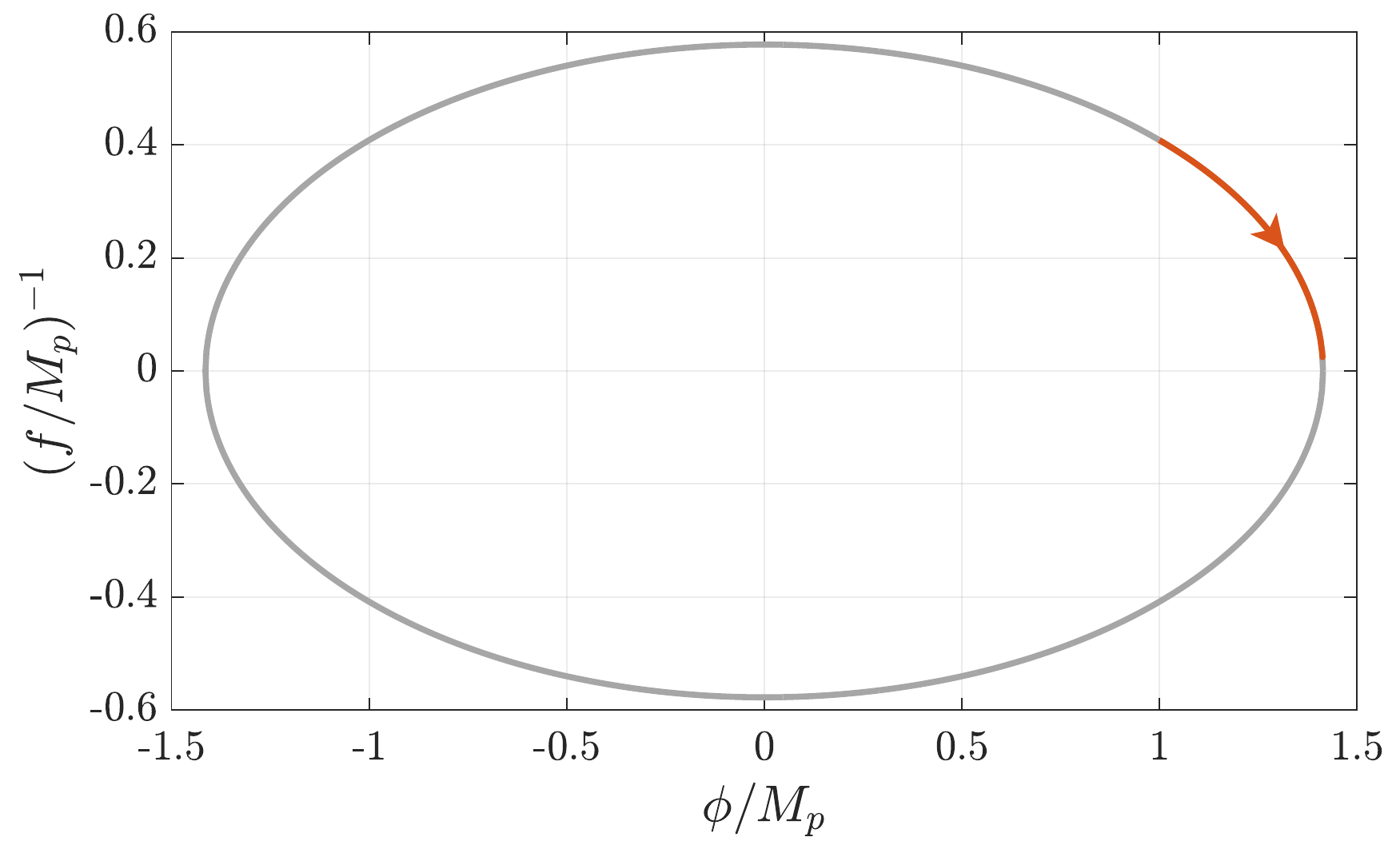}
\caption{Analytical equation of the ellipse in Eq.~(\ref{eq:K}) with $c_1=M_p^2$ (grey curve), and trajectory of the fields obtained numerically by solving the background equations (red curve). The arrow indicates the direction of motion. Numerical integration is carried out until the end of inflation.}
\label{fig:ellipse}
\end{figure}

Following Ref.~\cite{Rinaldi:2015uvu}, we consider the following classically scale-invariant action which, in the Jordan frame (hence the subscript $_J$), features a scalar field $\phi$ non-minimally coupled to gravity, and is given by:~\footnote{This action is in principle not the most general action quadratic in curvature invariants~\cite{Stelle:1976gc,Salvio:2018crh}, as it lacks a term quadratic in the Weyl tensor. The reason we have not included this is that we work in a FLRW background, which is conformally flat, and whose Weyl tensor therefore vanishes. Consequently, at the classical level, including or not a term squared in the Weyl tensor will not change our results. While this is true at the unperturbed level, even at the classical level the situation can be very different once perturbations around the classical background are introduced (see e.g.\ Ref.~\cite{DeFelice:2023psw} for a recent study). However, an investigation of this point goes well beyond the scope of our work.}
\begin{equation}
S_J = \int d^4 x \sqrt{-g} \left [ \dfrac{\alpha}{36}R^2 + \dfrac{\xi}{6}\phi^2 R -\dfrac{1}{2}\partial_{\mu}\phi\partial^{\mu}\phi -\dfrac{\lambda}{4}\phi^4 \right ] \,,
\label{eq:actionjf}
\end{equation}
where $\alpha$, $\xi$, and $\lambda$ are arbitrary dimensionless constants to be constrained through the analysis we will carry out in this work. We note that the JF action is the on-shell equivalent of:
\begin{equation}
S_J = \int d^4 x \sqrt{-g} \left [ \left ( \dfrac{\alpha\psi^2}{18} +\dfrac{\xi\phi^2}{6} \right ) R -\dfrac{\alpha\psi^4}{36} - \dfrac{1}{2}\partial_{\mu}\phi\partial^{\mu}\phi-\dfrac{\lambda}{4}\phi^4 \right ] \,.
\label{eq:actionjfonshell}
\end{equation}
The reason is that the equation of motion for the auxiliary field $\psi$ fixes $\psi^2=R$. Consequently, the results obtained by Ferreira \textit{et al.} in Ref.~\cite{Ferreira:2018qss} for a scale-invariant model with two scalar fields non-minimally coupled to gravity also apply (on-shell) here, see also Ref.~\cite{Garcia-Bellido:2011kqb} for a very important earlier work in the context of a scale-invariant Higgs-dilaton model.~\footnote{We emphasize that scale-invariance is a key feature of both our model and the Higgs-dilaton model~\cite{Trashorras:2016azl,Casas:2017wjh}. Indeed, such a feature is evident in the effective single-field dynamics. Moreover, in the Higgs-dilaton model, the non-minimal coupling to the dilaton field is consistent with zero -- as the analysis of Ref.~\cite{Trashorras:2016azl} shows, cosmological observations set a stringent upper limit on the non-minimal coupling of order $\sim 10^{-3}$, comparable to the bound we will obtain on $\xi$ later on in our analysis -- thereby recovering Starobinsky's results. However, a consistent comparison with Ref.~\cite{Trashorras:2016azl} is somewhat challenging, given that the model in question also accounts for a dark energy phase which, given its connection to the inflationary phase, imposes additional constraints on the spectral indices which are not present in our case.} By means of the following Weyl transformation:
\begin{equation}
\tilde{g}_{\mu \nu}=e^{2\omega(x)}g_{\mu \nu} \,,
\label{eq:weyltransformation}
\end{equation}
where we have defined
\begin{equation}
\omega \equiv \frac{1}{2}\ln \frac{2}{M^2} \left ( \dfrac{\alpha\psi^2}{18} +\dfrac{\xi\phi^2}{6} \right ) \,,
\label{eq:omega}
\end{equation}
one can move to the Einstein frame after introducing the field $\mathfrak{f}=Me^{-\omega}$. The action can then be written in compact form:
\begin{equation}
S_E = \int d^4 x \sqrt{-g} \left [ \dfrac{M^2}{2}R - \dfrac{1}{2}\mathcal{G}_{IJ}g^{\mu\nu}\partial_{\mu}\phi^I\partial_{\nu}\phi^J - V(\phi^I) \right ] \,,
\label{eq:efactionphif}
\end{equation}
where 
\begin{equation}
\phi^I \equiv
\begin{pmatrix}
\phi \\
\mathfrak{f}
\end{pmatrix}\,, \quad \quad
\mathcal{G}_{IJ} \equiv
\begin{pmatrix}
e^{2b(\mathfrak{f})} & 0\\
0 & 6e^{-2b(\mathfrak{f})}
\end{pmatrix}\,,
\label{eq:phiffieldspace}
\end{equation}
and $b(\mathfrak{f})$ is defined as:
\begin{equation}
b(\mathfrak{f}) \equiv \ln \left ( \dfrac{\mathfrak{f}}{M} \right ) \,.
\label{eq:b}   
\end{equation}
The potential in the Einstein frame appearing in Eq.~(\ref{eq:efactionphif}) is then given by:
\begin{equation}
V(\phi^I) = V \left ( \phi, \mathfrak{f} \right ) = -\dfrac{3\xi\phi^2\mathfrak{f}^2}{2\alpha} + \dfrac{\Omega \phi^4\mathfrak{f}^4}{4\alpha M^4} + \dfrac{9M^4}{4\alpha}\,,
\label{eq:potentialphif}
\end{equation}
where $\Omega$ has been defined as follows:
\begin{equation}
\Omega \equiv \alpha\lambda + \xi^2 \,.
\label{eq:omegaalphalambdaxi}
\end{equation}
In Eq.~(\ref{eq:potentialphif}), $M$ is an arbitrary parameter with mass dimension $1$. As emphasized in Ref.~\cite{Rinaldi:2015uvu}, $M$ is a redundant parameter of the theory -- in fact, variation of the action with respect to $M$ leads to an equation which is manifestly vanishing on-shell. Upon inspecting Eq.\eqref{eq:efactionphif} it is natural to identify $M$ with $M_p$, and we will adopt this identification from now on. We stress that the appearance of the parameter $M$ has nothing to do with the breaking of scale-invariance, which is preserved in the EF. For this reason, we can evaluate the Noether current associated to scale symmetry, $K_{\mu}$, given by (see also Ref.~\cite{Garcia-Bellido:2011kqb}):
\begin{equation}
K_{\mu} \equiv \partial_{\mu}K\,, \quad \quad K \equiv \dfrac{M_p^2}{2} \left ( \dfrac{\phi^2}{M_p^2} + \dfrac{6M_p^2}{\mathfrak{f}^2} \right ) \,,
\label{eq:K}
\end{equation}
which is covariantly conserved along the equations of motion, i.e.\ $\nabla_{\mu}K^{\mu} = 0$. The explicit solution to the conservation equation takes the following form:
\begin{equation}
K = c_1 + c_2 \int \dfrac{d t}{a^3(t)}\,.
\label{eq:ksolutionscalesymmetrybreaking}
\end{equation}
This shows that $K$ quickly approaches a constant value, thereby spontaneously breaking scale symmetry. Then, the motion in the $(\phi, \mathfrak{f}^{-1})$ plane is constrained to lie along an ellipse, as shown in Fig.~\ref{fig:ellipse} (see also the discussion in Refs.~\cite{Garcia-Bellido:2011kqb,Ferreira:2018qss}). Without loss of generality, we can set $c_1 = M_p^2$. Then, along the elliptic orbit, $\mathfrak{f}$ can always be expressed in terms of $\phi$ as:
\begin{equation}
\mathfrak{f} =  \dfrac{\sqrt{6}M_p^2}{\sqrt{2M_p^2-\phi^2}}\,.
\label{eq:fphi}
\end{equation}
It is well known that models of multi-field inflation can display a simplified behavior when a symmetry is at play, as a result of the conserved Noether current: it is the case for Higgs inflation, where the $SU(2)$ gauge symmetry manifests as an $SO(4)$ symmetry in fields space (see e.g.\ Ref.~\cite{Greenwood:2012aj}), but also for the Higgs-dilaton model of inflation (as well as dark energy, see e.g.\ Refs.~\cite{Trashorras:2016azl,Casas:2017wjh}) and scale-invariant generalizations thereof (see e.g.\ Ref.~\cite{Casas:2018fum}).

\section{Theoretical aspects}
\label{sec:theory}

We present the main analytical results concerning theoretical aspects of the model, which will later be useful in the numerical analysis. We begin in Sec.~\ref{subsec:entropy}, where we introduce the methodology adopted and prove that entropy perturbations vanish due to scale symmetry. In Sec.~\ref{subsec:redefinition}, we define auxiliary fields which simplify the analysis and compute all relevant quantities required for the later numerical implementation (see Sec.~\ref{sec:cosmoconstraints}). In Sec.~\ref{subsec:nonGaussianity}, we quantify the level of local non-Gaussianity predicted in the squeezed limit.

\subsection{Vanishing entropy perturbations}
\label{subsec:entropy}

As originally proposed in Refs.~\cite{Gordon:2000hv,DiMarco:2002eb} (see also Ref.~\cite{Achucarro:2010da}), to facilitate the interpretation of the evolution of cosmological perturbations, as well as their interrelation, we introduce an orthonormal basis in field space wherein various quantities dependent on the fields are decomposed into so-called adiabatic and entropy components. Considering the 2D field space defined in Eq.~(\ref{eq:phiffieldspace}), the unit vectors defining this basis are:
\begin{equation}
\textbf{u}_{\sigma}^I\equiv \frac{1}{\sqrt{e^{2b}\dot{\phi}^2+6e^{-2b}\dot{\mathfrak{f}}^2}} \left ( \dot{\phi}, \dot{\mathfrak{f}} \right ) \,, \quad \quad \textbf{u}_{s}^I\equiv \frac{1}{\sqrt{e^{2b}\dot{\phi}^2+6e^{-2b}\dot{\mathfrak{f}}^2}} \left ( -\sqrt{6}e^{-2b}\dot{\mathfrak{f}}, \dfrac{e^{2b}}{\sqrt{6}}\dot{\phi} \right ) \,,
\label{eq:ufields}
\end{equation}
where $\textbf{u}_{\sigma}^I$ is tangent to the background trajectory and $\textbf{u}_{s}^I$ is orthogonal to it by construction. It is indeed easy to show that:
\begin{equation}
\textbf{u}_{\sigma}^I\textbf{u}_{\sigma J}+\textbf{u}^I_{s}\textbf{u}_{s J}=\delta^I_J\,,
\label{eq:usigmaujdeltaij}
\end{equation}
and therefore
\begin{equation}
\dot{\sigma}^2\equiv e^{2b}\dot{\phi}^2+6e^{-2b}\dot{\mathfrak{f}}^2\,,
\label{eq:sigmadot}
\end{equation}
where the adiabatic field $\sigma$ is defined via:
\begin{equation}
d\sigma = e^b \cos\theta d\phi + \sqrt{6}e^{-b}\sin\theta d\mathfrak{f}\,.
\label{eq:sigma}
\end{equation}
We refer to the two orthonormal unit vectors $\textbf{u}_{\sigma}$ and $\textbf{u}_{s}$ as defining the adiabatic and entropy components respectively.

Following Refs.~\cite{Lalak:2007vi,DeAngelis:2023fdu}, we find it convenient to introduce the notation:
\begin{equation}
\textbf{u}_{\sigma}^I = \left ( e^{-b}\cos\theta, \dfrac{e^b}{\sqrt{6}}\sin\theta \right )\,, \quad \quad \textbf{u}_{s}^I= \left ( -e^{-b}\sin\theta, \dfrac{e^b}{\sqrt{6}}\cos\theta \right ) \,,
\label{eq:utrig}
\end{equation}
where $\theta$ is the rotation angle with respect to the tangent of the background trajectory. Within this formalism, adiabatic and entropy perturbations, which we denote by $\delta s$ and $\delta\sigma$, are defined as:
\begin{equation}
\delta s = \textbf{u}_{s I}\delta \phi^I\,, \quad \quad \delta \sigma = \textbf{u}_{\sigma I}\delta \phi^I\,.
\label{eq:adiabaticentropy}
\end{equation}
In particular, we can evaluate entropy perturbations imposing the constraint given by Eq.~(\ref{eq:fphi}):
\begin{equation}
\delta s = \sqrt{6} e^{-b} \cos\theta\delta \mathfrak{f} - e^b\sin\theta \delta \phi = \left [ \dfrac{12e^{-b}M_p^2\phi}{(2M_p^2-\phi^2)^{3/2}}\cos\theta - e^b\sin\theta \right ] \delta\phi=0\,,
\label{eq:deltas}
\end{equation}
where the last equality can easily be obtained by employing the explicit expressions for $\sin\theta$ and $\cos\theta$ and further writing $\dot{\mathfrak{f}} = (\partial \mathfrak{f}/\partial \phi) \Dot{\phi}$. This therefore proves that entropy perturbations vanish in our model, as a consequence of the constraint given by Eq.~(\ref{eq:fphi}). This, in turn, follows from conservation of the Noether current associated to scale symmetry. Therefore, the absence of entropy perturbations is ultimately a consequence of scale-invariance. As a side note, this result rules out any concern about a tachyonic mass of entropy perturbations. This fact is known to plague several multi-field inflationary models with hyperbolic geometry, as discussed in Ref.~\cite{Renaux-Petel:2015mga}, and eventually leads to inflation ending prematurely. In this regard, scale-invariance protects from any form of geometrical destabilization. At the same time, the formalism we have adopted is safe from the apparent destabilization effects investigated by Cicoli \textit{et al.} in Ref.~\cite{Cicoli:2021yhb}. The authors stress the importance of properly defining the entropy variable before claiming the presence of growing isocurvature perturbations. In light of the result obtained in Eq.~\eqref{eq:deltas}, the decomposition in tangent and normal perturbations with respect to the inflationary trajectory adopted here is free from ambiguities. 

\subsection{Field redefinition and observable predictions}
\label{subsec:redefinition}

In what follows we show that our model can be treated in the same vein as single-field inflation, given that its dynamical content can be shifted to one field out of the two, with the other one behaving effectively as a spectator field. To do so, we further exploit the constraint in Eq.~(\ref{eq:K}) by defining the new fields $\rho$ and $\chi$ (see e.g.\ Ref.~\cite{Tambalo:2016eqr}):
\begin{eqnarray}
\label{eq:rho}& \rho &= \sqrt{6}M_p \arcsinh\left(\dfrac{\phi \mathfrak{f}}{\sqrt{6}M_p^2}\right), \\
\label{eq:chi}& \chi &= \dfrac{M_p}{2}\ln \left(\dfrac{\phi^2}{2M_p^2} + \dfrac{3M_p^2}{\mathfrak{f}^2}\right).
\end{eqnarray}
The EF action in the $(\phi, \mathfrak{f})$ representation given in Eq.~(\ref{eq:efactionphif}) can then be written in the $(\rho, \chi)$ representation in the following compact form:
\begin{equation}
S_E =  \int d^4 x \sqrt{-g} \left [ \dfrac{M_p^2}{2}R - \dfrac{1}{2}\mathcal{G}_{IJ}g^{\mu\nu}\partial_{\mu}\phi^I\partial_{\nu}\phi^J - V(\phi^I) \right ] \,,
\label{eq:efactionrhochi}
\end{equation}
where:
\begin{equation}
\phi^I \equiv
\begin{pmatrix}
\rho \\
\chi
\end{pmatrix}\,, \quad \quad
\mathcal{G}_{IJ} \equiv
\begin{pmatrix}
1 & 0\\
0 & e^{2b(\rho)}
\end{pmatrix}\,,
\end{equation}
and $b(\rho)$ is defined as:
\begin{equation}
b(\rho) = \dfrac{1}{2}\ln \left [ 6\cosh^2 \left ( \dfrac{\rho}{\sqrt{6}M_p} \right ) \right ] \,.
\label{eq:br}
\end{equation}
The potential in this field representation depends exclusively on $\rho$, and takes the form:~\footnote{The potential takes the form of a Mexican hat, whose minimum however is non-vanishing. As a result, at the stable fixed point, there is in principle a residual cosmological constant. In Ref.~\cite{Rinaldi:2015uvu} it was shown that a specific combination of the couplings can lead to the $R^2$ and $\phi^4$ terms exactly canceling this residual cosmological constant, at the price of the resulting spectral indices not agreeing with observations: for observationally allowed values of the parameters, as per our later analysis, the residual cosmological constant would be very large, and a cancellation mechanism would therefore be required. A potential mechanism in this sense was studied in Ref.~\cite{vandeBruck:2021xkm}, and invokes the introduction of a third field, which still respects the underlying scale symmetry, but at the same time contributes to the inflationary dynamics (i.e.\ it is not a spectator field), thereby affecting the model's observational predictions. It would be interesting to study such an extension using the same method adopted in this paper, and we leave this very interesting study to follow-up work.}
\begin{eqnarray}
V(\phi^I) = V(\rho) = \dfrac{9M_p^4}{4\alpha}\left[1-4\xi \sinh^2 \left ( \dfrac{\rho}{\sqrt{6}M_p} \right )+4\Omega \sinh^4 \left ( \dfrac{\rho}{\sqrt{6}M_p} \right ) \right]\,.
\label{eq:potentialrhochi}
\end{eqnarray}
At the background level, the homogeneous Klein-Gordon equations for the two fields are:
\begin{eqnarray}
\label{eq:KGrho}&\Ddot{\rho}&+3H\Dot{\rho}+V_{,\rho} = b_{,\rho}\,e^{2b(\rho)}\Dot{\chi}^2\,,\\
\label{eq:KGchi}&\Ddot{\chi}&+3H\Dot{\chi}+2b_{,\rho}\,\Dot{\chi}\Dot{\rho}=0\,.
\end{eqnarray}
The Einstein equations determining the evolution of the scale factor are:
\begin{eqnarray}
\label{eq:Einstein1}
&H^2&=\dfrac{1}{3M_p^2} \left ( \dfrac{\Dot{\rho}^2}{2}+e^{2b(\rho)}\dfrac{\Dot{\chi}^2}{2} + V(\rho) \right ) \,,\\
\label{eq:Einstein2}
&\Dot{H}&= -\dfrac{1}{2M_p^2}\left(\Dot{\rho}^2+e^{2b(\rho)}\Dot{\chi}^2\right)\,.
\end{eqnarray}
From the above equations it is clear that $\chi$ plays the role of the Goldstone boson of the theory: when scale symmetry is spontaneously broken, $K$ approaches a constant and consequently so does $\chi$. The relevant degree of freedom for the inflationary dynamics is then $\rho$, consistently with the fact that $V(\phi^I) = V(\rho)$. Indeed, considering the slow-roll approximation, from Eq.~(\ref{eq:KGchi}) it is easy to see that the dynamics are the ones of single-field inflation.

The above results translate into having $\theta \approx 0$ once Eqs.~(\ref{eq:ufields},\ref{eq:utrig}) for the new field representation are employed. If we introduce the convenient notation $V_{,s} = \textbf{u}^I_{s}V_{,I}$, the contribution of the entropy field given via:
\begin{equation}
\dot{\theta}= -\dfrac{V_{,s}}{\Dot{\sigma}} -b_{,\rho}\, \Dot{\sigma}\sin\theta\,,
\label{eq:dottheta}
\end{equation}
is therefore identically zero, implying that the background trajectory is flat. Perturbations along the orthogonal direction are not coupled through the potential either (since $V_{,s} = 0$). As a consequence, the entire dynamics of the system are controlled by the tangent field $\sigma$ which, in the slow-roll approximation, is governed by the relation:~\footnote{Here and in what follows, the symbol $\simeq$ indicates that we are within the slow-roll approximation.}
\begin{equation}
\dfrac{\Ddot{\sigma}}{H\Dot{\sigma}}\simeq \epsilon -\dfrac{V_{,\rho\rho}}{3H^2}\,.
\label{eq:ddotsigmaapprox}
\end{equation}
Moreover, the primordial scalar power spectrum at horizon crossing reads:
\begin{equation}
\mathcal{P}_{\zeta}(k)\simeq \frac{H^2}{8\pi^2 \,\epsilon}\biggl|_{k=k_*}\,,
\label{eq:primordial}
\end{equation}
and following Ref.~\cite{DeAngelis:2023fdu} we can therefore obtain the spectral parameters (at the scale $k=k_*$). In particular, we find that the spectral index is given by:
\begin{equation}
n_s \simeq -6\epsilon + \dfrac{2V_{,\rho\rho}}{3H^2}\,,
\label{eq:ns}
\end{equation}
whereas the running of the spectral index $\alpha_s \equiv dn_s/d\ln k$ takes the form:
\begin{equation}
\alpha_s \simeq -24\epsilon^2+16\epsilon\dfrac{V_{,\rho\rho}}{3H^2} + 2\sqrt{2\epsilon}\cos\theta\dfrac{V_{,\rho\rho\rho}}{3H^2}\,.
\label{eq:alphas}
\end{equation}
For completeness, even though it is not used in the later numerical analysis, we quote the expression for the running of the running of the scalar spectral index $\beta_s \equiv d\alpha_s/d\ln k$:
\begin{align}
\beta_s \simeq& -192 \epsilon^3 + \dfrac{64\epsilon^2V_{, \rho\rho}}{H^2} - \dfrac{32\epsilon V_{, \rho\rho}^2}{9H^4} + \dfrac{8\sqrt{2}\epsilon^{3/2}\cos\theta V_{,\rho\rho\rho}}{H^2}-\dfrac{2\sqrt{2\epsilon}\cos\theta V_{,\rho\rho}V_{,\rho\rho\rho}}{9H^4}\nonumber \\&+\dfrac{4\epsilon\cos^2\theta V_{,\rho\rho\rho\rho}}{3H^2}\,.
\end{align}
The transfer matrix formalism~\cite{Wands:2002bn} allows us to show that the effects of isocurvature modes on the adiabatic ones are absent on super-horizon scales. Indeed, the transfer functions which relate the power spectrum at the end of inflation to the power spectrum at horizon crossing are given by:
\begin{equation}
\begin{aligned}
\mathcal{T}_{\zeta \mathcal{S}}(t_*,t)&=\int^t_{t_*} A(t')H(t')\mathcal{T}_{\mathcal{S} \mathcal{S}}(t_*,t')dt',\\
\mathcal{T}_{\mathcal{S} \mathcal{S}}(t_*,t')&=\text{exp}\left(\int^{t'}_{t_{*}} B(t'')H(t'')dt''\right),
\label{eq:transf}
\end{aligned}
\end{equation}
where the time-dependent dimensionless functions $A$ and $B$ are given by~\cite{DeAngelis:2023fdu}:
\begin{align}
& A = 2\xi_1 \sin\theta \simeq 0\,, \label{eq:dimensionlessa} \\
& B = -2\epsilon +\xi_1 + \dfrac{\xi_1^2}{3} + \dfrac{\xi_2}{3} + \dfrac{V_{,\rho\rho}}{3H^2}\,, \label{eq:dimensionlessb}
\end{align}
with $\xi_1 \equiv \sqrt{2\epsilon} b_{,\rho}$ and $\xi_2 \equiv 2\epsilon b_{,\rho\rho}$. Hence, by means of Eq.~(\ref{eq:dimensionlessa}) and the fact that $\theta \approx 0$, we see that the power spectrum at the end of inflation does not deviate from the one at horizon crossing, since the two are related by a factor $(1+\mathcal{T}_{\zeta \mathcal{S}}^2)$.

\subsection{Non-Gaussianity}
\label{subsec:nonGaussianity}

Primordial cosmological perturbations are usually expressed in terms of the curvature perturbation on uniform energy density hypersurfaces, which reads:
\begin{equation}
\zeta = \Phi -\dfrac{H}{\Dot{H}}\left(\Dot{\Phi} + H\Phi\right) \simeq H \left ( \dfrac{\Dot{\rho}\delta\rho + e^{2b(\rho)}\Dot{\chi}\,\delta \chi}{\Dot{\rho}^2 + e^{2b(\rho)}\Dot{\chi}^2} \right ) \,.
\label{eq:zeta}
\end{equation}
where the second equality is obtained by considering the equations of motion at first order in perturbations, and is valid only within the long-wavelength and slow-roll limits. 
Moreover, this same quantity evaluated on some final uniform-density spacetime slice $t_c$ can be expressed in terms of spatial fluctuations in $e$-folding number between an initially flat slice (at $t=t_*$) and the final comoving one (at $t=t_c$):
\begin{equation}
\zeta(t_c,\textbf{x})\simeq \delta N (t_c,t_*,\textbf{x})\,,
\label{eq:zetadelta}
\end{equation}
allowing us to compute the non-linear evolution of cosmological perturbations on large scales without the need to solve the full perturbed field equations. In line with the so-called $\delta N$ expansion for the field perturbation~\cite{Sasaki:1995aw,Sugiyama:2012tj}, we can write:
\begin{equation}
\delta N(t_c,t_*, \mathbf{x}) = N_{,I}(N, \phi^J_*)\delta \phi^I_*(\mathbf{x})  +\dfrac{1}{2}N_{,IJ}(\phi^K_*)\delta \phi^I_*(\mathbf{x})\delta \phi^J_*(\mathbf{x}) + \dots\,
\label{eq:deltan}
\end{equation}
retaining only terms up to second order. For sufficiently small $\delta \phi_*^I(\mathbf{x}) \equiv \phi_*^I(\mathbf{x}) - \phi_*^I$, the two points $\phi_*^I(\mathbf{x})$ and $\phi_*^I$ are connected by a unique geodesic that can be parametrized by $\lambda$. By introducing the quantity $\mathcal{Q}^I = d\phi^I/ d\lambda|_{\lambda=0}$ which resides in the tangent space at $\phi^I(\lambda = 0)$ and transforms covariantly \cite{Mori:2017caa,Gong:2011uw}, we can express $\delta \phi^I$ in terms of $\mathcal{Q}^I$ as:
\begin{equation}
\delta \phi^I = \mathcal{Q}^I - \dfrac{1}{2!}\Gamma^{I}_{JK}\mathcal{Q}^J\mathcal{Q}^K+ \dots\,,
\label{eq:deltaphii}
\end{equation}
from which Eq.~(\ref{eq:deltan}) can be recast as follows:
\begin{equation}
\zeta(N, \mathbf{x}) = N_{, I}(N, \phi^J_*)\mathcal{Q}^I_*(\mathbf{x}) + \dfrac{1}{2}\mathcal{D}_I\mathcal{D}_J N(N, \phi^K_*)\mathcal{Q}^I_*(\mathbf{x})\mathcal{Q}^J_*(\mathbf{x}) + \dots\,,
\label{eq:covariant}
\end{equation}
where once more we have retained only terms up to second order.

We now turn our attention to correlation functions of the curvature perturbation. Moving to Fourier space, we parametrize the two-point correlation function as:
\begin{equation}
\langle \zeta(\mathbf{k}_1)\zeta(\mathbf{k}_2)\rangle = (2\pi)^3\delta^3(\mathbf{k}_1 + \mathbf{k}_2 )P_{\zeta}(k_1) = (2\pi)^3\delta(\mathbf{k}_1 + \mathbf{k}_2)\dfrac{2\pi^2}{k_1^3}\mathcal{P}_{\zeta}(k_1)\,,
\label{eq:twopoint}
\end{equation}
and similarly for the three-point correlation function:
\begin{equation}
\langle \zeta(\mathbf{k}_1)\zeta(\mathbf{k}_3)\zeta(\mathbf{k}_3) \rangle = (2\pi)^3\delta^3(\mathbf{k}_1 + \mathbf{k}_2 + \mathbf{k}_3)B_{\zeta}(k_1, k_2, k_3)\,,
\label{eq:threepoint}
\end{equation}
where $P_{\zeta}(k)$ and $\mathcal{P}_{\zeta}(k)$ are respectively the power spectrum and reduced power spectrum, whereas $B_{\zeta}(k_1, k_2, k_3)$ is the bispectrum. To quantify the level of non-Gaussianity we introduce the parameter $f_{\text{NL}}$:
\begin{equation}
f_{\text{NL}} = \dfrac{5}{6}\dfrac{B_{\zeta}(k_1, k_2, k_3)}{P_{\zeta}(k_1) P_{\zeta}(k_2) + \mathrm{c.p.}}\,,
\label{eq:fnl}
\end{equation}
where c.p.\ denotes cyclic permutations of $k_1, k_2$ and $k_3$. As a side note, we stress that the $\delta N$ formalism is only valid in the squeezed limit of the three-point correlation function. From Eq.~(\ref{eq:covariant}) and following Ref.~\cite{Mori:2017caa}, we find that the power spectrum is given by:
\begin{equation}
\mathcal{P}_{\zeta} = \left ( \dfrac{H_*}{2\pi} \right ) ^2\mathcal{G}_*^{IJ}N_{,I}N_{,J} = \left ( \dfrac{H_*}{2\pi}\right ) ^2 (N_{,\rho})^2\,,
\label{eq:powerspectrum}
\end{equation}
where $N\simeq \int d\rho/(M_p \sqrt{2\epsilon(\rho)})$. On the other hand, working within the slow-roll approximation, we find that $f_{\text{NL}}$ is given by:
\begin{equation}
f_{\text{NL}} = \dfrac{5}{6}\dfrac{\mathcal{G}_*^{IK} N_{,K}\mathcal{G}_*^{JL} N_{,L} \mathcal{D}_I N_{,J}}{(\mathcal{G}_*^{KH}N_{,H}N_{,K})^2} = \dfrac{5}{6}\dfrac{N_{,\rho}^2 (N_{,\rho\rho} -\Gamma^{\rho}_{\rho\rho}N_{,\rho})}{N_{,\rho}^4} = \dfrac{5}{6}\dfrac{N_{,\rho\rho}}{N^2_{,\rho}} \simeq \dfrac{5}{3} M \epsilon(\rho)\dfrac{\partial}{\partial \rho}\dfrac{1}{\sqrt{2\epsilon(\rho)}}\,.
\end{equation}
The above expression is identical to the one obtained for single-field inflation models with canonical kinetic terms, see e.g.\ Eq.~(1.34) of Ref.~\cite{Byrnes:2014pja}. This result is indeed consistent with the fact, already stressed earlier, that the inflationary dynamics within our model are effectively driven by the field $\rho$, whose kinetic term is canonical.

Given the slow-roll suppression of $f_{\text{NL}}$, we expect our model to be in excellent agreement with current upper limits on the amount of non-Gaussianity. At the same time, it is meaningful to account for the -- equally small -- contribution of non-Gaussianity of the field perturbations at horizon crossing. By making use of the Maldacena consistency relation, which is naturally embedded in the $\delta N$ formalism~\cite{Abolhasani:2018gyz}, we find~\cite{Byrnes:2014pja}:
\begin{equation}
f_{\text{NL}} = \dfrac{5}{12}(1-n_s)\,.
\label{eq:fnlns}
\end{equation}
We have explicitly verified that, within the parameter space constrained through the numerical analysis to be discussed later in Sec.~\ref{sec:cosmoconstraints}, $f_{\text{NL}}$ is always well within the upper limits on the amount of non-Gaussianity obtained from the temperature and polarization CMB bispectra measured by the \textit{Planck} satellite. To be more concrete, we consider limits on the amplitude of the local bispectrum, $f_{\text{NL}}^{\text{local}} = -0.9 \pm 5.1$~\cite{Planck:2019kim}. We focus on the local template for three reasons: \textit{i)} it peaks in the squeezed limit, which is also the limit within which the $\delta N$ formalism allows us to compute the amplitude of the bispectrum, \textit{ii)} it is the one which can be used to discriminate single-field from multi-field inflation, and \textit{iii)} its amplitude is the most tightly constrained among the standard templates considered (local, equilateral, orthogonal). Then, purely by means of example, for the benchmark point in parameter space given by $\xi = 0.00039$, $\alpha = 1.87 \times 10^{10}$, $\Omega = 1.64 \times 10^{-7}$, and $\rho_{\star} = 5.15\,M_p$, with the subscript $_{\star}$ denoting a quantity evaluated at horizon crossing, we have $f_{\text{NL}}^{\text{local}}=-0.015$, well within the region allowed by present precision cosmological observations.~\footnote{Note that constraints on $f_{\text{NL}}^{\text{local}}$ from large-scale structure probes are weaker by at least an order of magnitude, see for instance Refs.~\cite{Cabass:2022wjy,DAmico:2022gki,Cabass:2022ymb,Barreira:2022sey,Cagliari:2023mkq}.}

\section{Cosmological Constraints}
\label{sec:cosmoconstraints}

We are now in the position to test scale-invariant inflation against precision cosmological data. We begin in Sec.~\ref{subsec:methods} by reviewing the numerical methodology adopted in the analysis. In Sec.~\ref{subsec:results}, we then derive constraints on the model's free parameters from current observations. Subsequently, in Sec.~\ref{subsec:discussion} we compare the model's observational predictions with those of other benchmark scenarios, such as Starobinsky inflation, and discuss the implications for forthcoming experiments.

\subsection{Methods}
\label{subsec:methods}

Within the theoretical parametrization described in Sec.~\ref{subsec:redefinition}, our scale-invariant inflationary model reduces to a scenario where two scalar fields coexist in a non-trivial field space $\mathcal{G}_{IJ}(\rho,\chi)$. In total, we are left with three free parameters: $\xi$, $\Omega$, and $\alpha$. Using Monte Carlo (MC) techniques, we aim to explore the 3D parameter space of the model to compare its theoretical predictions against observational data. To do so, we strictly follow the methodology introduced by some of us in Ref.~\cite{Giare:2023kiv}, employing a sampling algorithm able to explore a large volume of parameter space and identify a sub-region where the model's predictions agree well with observations. Specifically, our algorithm works as follows:
\begin{enumerate}
\item At each step of the MC, we randomly select initial conditions for the fields and values for the three free parameters of the model from the uniform prior ranges reported in the rightmost column of Tab.~\ref{tab:results}. This therefore also allows us to explore the impact of initial conditions for the fields.
\item For the particular set of parameter values and initial conditions fixed at point 1., we integrate the equations of motion for the fields and track their evolution over the inflationary potential. Given the peculiarity of this model, where, depending on the parameter values, the field evolution can be characterized by a very prolonged slow-roll phase eventually leading to eternal inflation, the integration process is carried out for a (fairly large) maximum number of e-folds, set to $N_{\max}=10^7$. Throughout the integration, we dynamically calculate the slow-roll parameter $\epsilon$ until the condition $\epsilon=1$ is satisfied. If this condition is not met within $N_{\rm{max}}$ e-folds, the model is classified as eternal inflation and rejected. Conversely, if the condition is satisfied during the integration, the point at which $\epsilon=1$ in the parameter trajectory is considered as potential ending point for inflation. To confirm that the point in question represents the actual end of inflation, we conduct a multitude of tests detailed in Ref.~\cite{Giare:2023kiv}. Among other things, we check that \textit{i)} the fields are not active enough to initiate a second stage of inflation, \textit{ii)} the parameter $\epsilon$ remains $\epsilon \ne 0$ throughout the full evolution, and \textit{iii)} inflation lasts for a total number of e-folds $\Delta N > 70$ to account for the observed homogeneity and isotropy of the Universe. If the model satisfies all these conditions, we move to point 3., else we return to point 1.\ and select a new point in parameter space.
\item After ensuring that the model satisfies all the requirements discussed in the previous point, we reconstruct the full field dynamics during the entire inflationary phase, including how the slow-roll parameters and observables evolve as a function of $N$. By doing so, we can obtain the values of all slow-roll parameters at horizon crossing (set to $N_{\star}=55$ e-folds before the end of inflation) including the amplitude of the primordial power spectrum of scalar perturbations ($A_s$), its spectral index ($n_s$), the running of the spectral index ($\alpha_s$), and the amplitude of tensor perturbations (characterized by the tensor-to-scalar ratio $r \equiv A_t/A_s$).~\footnote{A precise estimate of $N_{\star}$ would require an analysis of the reheating stage, which is beyond the scope of the present work. Nevertheless, we are confident about our choice of fixing $N_{\star}=55$ in the Einstein frame given that, as we will discuss below in Eq.~(\ref{eq:nsrstarobinsky}), the same relation between $n_s$ and $r$ known for Starobinsky's model holds true in this model. Stated differently, the spectral indices and the number of e-folds are closely related by $n_s\simeq 1-2/N_{\star}$ and $r\simeq 12/N_{\star}^2$. For this reason, we do not expect significant deviations from the preferred value in Starobinsky's inflation, usually set to $N_{\star}=55$, with changes of less than one \textit{e}-fold when the reheating phase is also accounted for, see Ref.~\cite{Bezrukov:2011gp}. We defer a full study of the reheating phase, expanding further along the lines of Ref.~\cite{Rinaldi:2015uvu}, to future work.} Additionally, our code is able to take into account the super-horizon evolution and transfer of entropy between isocurvature and scalar perturbations by means of the transfer matrix formalism detailed in Refs.~\cite{DeAngelis:2023fdu,Giare:2023kiv}. As argued on the basis of analytical considerations in Sec.~\ref{subsec:entropy}, scale-invariance prevents the fields from transferring entropy perturbations. Interestingly, we find that the entropy transfer between isocurvature and scalar perturbations is consistent with zero also at the numerical level, confirming the theoretical result. A direct consequence of this fact is that the observational values for the inflationary parameters are set at horizon crossing (just as in single-field models) and remain unchanged at the end of inflation. To state it differently, the power spectrum at the end of inflation is equal to the one at horizon crossing, given that $\mathcal{T}_{\zeta \mathcal{S}}=0$ and thus $(1+\mathcal{T}_{\zeta \mathcal{S}}^2)=1$ [see Eq.~(\ref{eq:transf})].
\item We save the model predictions for observable quantities in a chain of points very similar to the output obtained by typical Markov Chains MC (MCMC) methods. We assign to each point in the chain a likelihood value obtained from an analytical multi-dimensional normal distribution:
\begin{equation}
\mathcal{L} \propto 
\exp\left(-\frac{1}{2} \left ( \boldsymbol{x}-\boldsymbol{\mu} \right ) ^{T} \boldsymbol{\Sigma}^{-1} \left ( \boldsymbol{x}-\boldsymbol{\mu} \right ) \right) \,,
\label{eq:like}
\end{equation}
where $\boldsymbol{\mu}$ and $\boldsymbol{\Sigma}$ represent the mean values and covariance matrix for the parameter vector $\boldsymbol{x} \equiv (A_s,n_s,\alpha_s,r)$ obtained from a joint analysis of the \textit{Planck} 2018 temperature and polarization (TT, TE, EE) and lensing reconstruction likelihoods~\cite{Planck:2018lbu,Planck:2019nip}, combined with the latest foreground-cleaned CMB B-mode power spectrum likelihood released by the \textit{BICEP/Keck} collaboration, based on observations from the \textit{BICEP2}, \textit{Keck Array}, and \textit{BICEP3} experiments up to and including the 2018 observation season~\cite{BICEP:2021xfz}.~\footnote{The chains used to determine the mean vector and covariance matrix we adopt have been obtained within the 9-parameter $\Lambda$CDM$+n_s+\alpha_s+\beta_s$ cosmological model~\cite{Giare:2023kiv}. Note that, in addition to Ref.~\cite{Giare:2023kiv}, a number of other earlier works have adopted and validated a similar compressed likelihood approach to constrain fundamental parameters from inflation-related observables, see e.g.\ Refs.~\cite{Visinelli:2018utg,Benetti:2021uea,Vagnozzi:2023lwo,Ben-Dayan:2023lwd,Winkler:2024olr}.} For further details on the validation of the method, we refer the reader to Ref.~\cite{Giare:2023kiv}.
\item We return to point 1.\ and keep sampling the model parameters.
\end{enumerate}
Using this method, we collect over $7 \times 10^{4}$ points within chains, each weighted by its own likelihood. This enables us to derive constraints on the free parameters of the model and study correlations both between the latter, as well as among observable quantities such as the spectral index and the amplitude of the primordial scalar and tensor power spectra, which of course are treated as derived parameters. We stress that, despite the similarity, our sampling method should \textit{not} be considered an MCMC algorithm (although it is able to recover the same results obtained using traditional MCMC methods, as discussed in detail throughout Ref.~\cite{Giare:2023kiv}). For instance, the sampling is completely random and there is no need to specify a proposal density or acceptance ratio (compare this to the widely used Metropolis-Hastings algorithm), thus the importance weight/multiplicity of each point in the chain is $1$ by construction. For the same reason, there is no in-built notion of convergence (\`{a} la Gelman and Rubin~\cite{Gelman:1992zz}), but the level of convergence is gauged empirically by assessing the stability of the resulting constraints against the addition of further samples.~\footnote{As a further empirical test of convergence, we verified that by accumulating an increasing number of points in the chains and marginalizing to find the 1D posterior distributions, the tails corresponding to 5\%~C.L. actually contain approximately 5\% of the total models. Such a result is exactly what is expected when marginalizing over the parameter space and allows us to safely conclude that the tails of the distributions were well-sampled and adequately populated.}

\subsection{Parameter Constraints}
\label{subsec:results}

\begin{figure}
\centering
\includegraphics[width=0.85\textwidth]{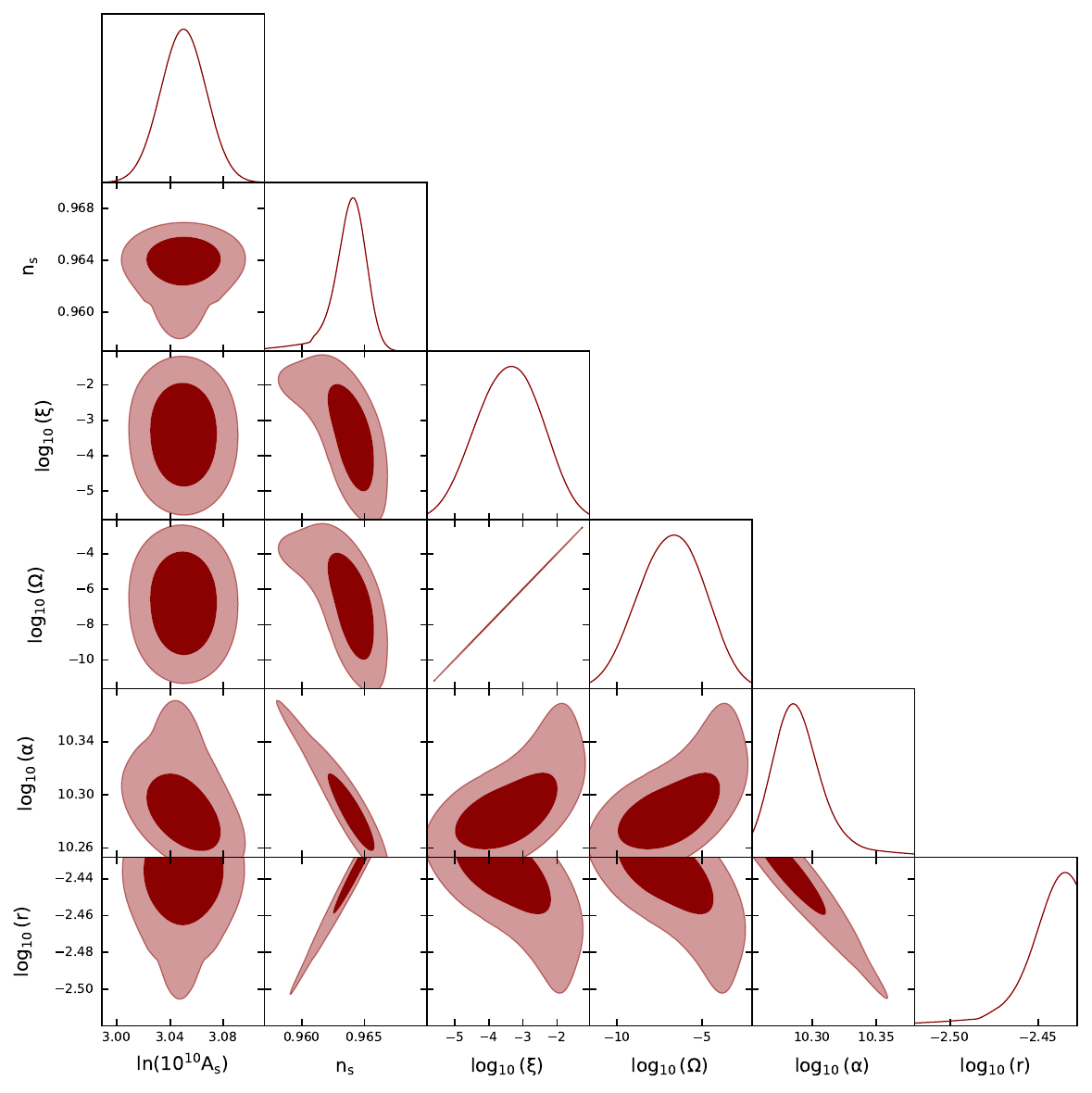}
\caption{Triangular plot showing 2D joint and 1D marginalized posterior probability distributions for a selection of parameters: the (natural logarithm of the) amplitude of primordial scalar perturbations $A_s$, the scalar spectral index $n_s$, the (logarithm of the) coupling parameters $\xi$, $\Omega$, and $\alpha$, with $\xi$ and $\alpha$ controlling the strength of the $\phi^2R$ and $R^2$ terms in the Jordan frame action respectively [see Eq.~(\ref{eq:actionjf})], and $\Omega/\alpha$ controlling the strength of the $\sinh^4(\rho)$ term in the potential within the $(\rho,\chi)$ field representation [see Eq.~(\ref{eq:potentialrhochi})], and the (logarithm of the) tensor-to-scalar ratio $r$.}
\label{fig:triangularplot}
\end{figure}

Constraints on the model parameters (including the initial conditions for the fields $\rho$ and $\chi$), as well as on the inflationary observables related to the primordial scalar and tensor power spectra (which are treated as derived parameters) are reported in Tab.~\ref{tab:results}. For two-tailed bounds we report 68\% confidence level (C.L.) intervals, whereas 95\% C.L. upper/lower limits are quoted for parameters whose distributions are not consistent with a ``detection''. In Fig.~\ref{fig:triangularplot} we instead show 2D joint and 1D marginalized posterior probability distributions for selected parameters (leaving out the distributions for the initial conditions).

Our first important finding is that, in spite of our choice of varying the initial conditions for the fields $\rho_{\text{ini}}$ and $\chi_{\text{ini}}$ within the flat priors reported in Tab.~\ref{tab:results}, these two parameters are entirely unconstrained. This emphasizes the fact that our model is not (or at least only weakly) sensitive to the choice of initial conditions, which is of course a positive aspect. In contrast, when focusing on the three free parameters of the model -- $\alpha$, $\xi$, and $\Omega$ (note that we impose a prior which is flat in $\log_{10}\xi$, rather than in $\xi$ itself) -- we observe that these are well constrained within the ranges set by our priors, with the bounds (and in particular the 68\%~C.L. intervals and/or 95\%~C.L. upper/lower limits) remaining well away from the upper and lower limits of the prior ranges. This indicates that the choice of prior ranges has virtually no effect on our constraints, and that the latter can be attributed to genuine physical effects of these parameters on observable quantities. To remind the reader about the meaning of these parameters, we recall that $\xi$ and $\alpha$ control the strength of the $\phi^2R$ and $R^2$ terms in the Jordan frame action respectively [see Eq.~(\ref{eq:actionjf})], whereas $\Omega/\alpha$ controls the strength of the $\sinh^4(\rho)$ term in the potential within the $(\rho,\chi)$ field representation [see Eq.~(\ref{eq:potentialrhochi})] -- with $\alpha$ and $\xi$ known, $\Omega$ itself is related to $\lambda$, which controls the strength of the quartic term in the Jordan frame action [see Eq.~(\ref{eq:actionjf})].

\begin{table}[!tb]
\begin{center}
\renewcommand{\arraystretch}{1.5}
\begin{tabular}{l c c}
\hline \hline
\textbf{Initial conditions} & \textbf{Constraints} & \textbf{Uniform prior ranges} \\
\hline
$\rho_{\rm ini}/ M_p$ & (\textit{unconstrained}) & $\rho_{\rm ini}/ M_{p} \in [0.1,2]$ \\
$\chi_{\rm ini}/ M_p$ & (\textit{unconstrained})  & $\chi_{\rm ini}/ M_{p} \in [0.1,10]$ \\
\hline \hline
\textbf{Model parameters}  & \textbf{Constraints} & \textbf{Uniform prior ranges} \\
\hline
$\xi$ & $<0.00142$ & $\log_{10}(\xi) \in [-5,-1]$ \\
$\alpha$ & $1.951^{+0.076}_{-0.11}\times 10^{10}$ & $10^{-10}\times\alpha \in [1,3]$\\
$\Omega$ & $0.93^{+0.72}_{-2.8} \times 10^{-5}$ & $\Omega\in [\xi^2,2\xi^2]$\\
\hline \hline
\textbf{Primordial spectra parameters}  & \textbf{Constraints} &  \\
\hline
$A_{s}$ & $\left(\,2.112\pm 0.033\,\right)\cdot 10^{-9}$ & (\textit{derived}) \\
$n_s$ & $0.9638^{+0.0015}_{-0.0010}$ & (\textit{derived}) \\
$\alpha_s$ & $<1.2 \times 10^{-4} $& (\textit{derived}) \\
$r$ & $> 0.00332$ & (\textit{derived}) \\
\hline \hline
\end{tabular}
\end{center}
\caption{External priors and observational constraints on the initial conditions of the fields $\rho$ and $\chi$ (first two rows, with $M_p$ denoting the Planck mass), the model parameters $\xi$ (more precisely $\log_{10}\xi$), $\alpha$, and $\Omega$ (three intermediate rows), and (derived) inflationary parameters controlling the primordial scalar and tensor power spectra $A_s$, $n_s$, $\alpha_s$, and $r$ (lower four rows). For what concerns observational constraints, for two-tailed bounds we report $1\sigma$ ($68\%$~C.L.) intervals, whereas for all other cases we report $2\sigma$ ($95\%$~C.L.) upper/lower bounds.}
\label{tab:results}
\end{table}

More in detail, the parameter $\alpha$ (controlling the strength of the $R^2$ term) is directly linked to the amplitude of the inflationary potential through Eq.~(\ref{eq:potentialrhochi}). It therefore affects the amplitudes of the primordial scalar and tensor power spectra. This connection is most evident if one observes the mutual correlations between $\alpha$, $A_s$, and $r$ in Fig.~\ref{fig:triangularplot}. Given that data from the \textit{Planck} satellite has been used to infer $A_s$ to high accuracy, through its effect on the amplitude of the acoustic peaks in the temperature and polarization anisotropy power spectra, constraints on $A_s$ usually impose stringent requirements on the amplitude of inflationary potentials, and therefore serve as calibrators for inflationary models. This holds true in our scale-invariant inflationary model as well, for which matching the amplitude of the primordial scalar power spectrum $A_s= (2.112\pm 0.033) \times 10^{-9}$, in excellent agreement with results in the literature, leads to the constraint $\alpha= 1.951^{+0.076}_{-0.11} \times 10^{10}$.

\begin{figure}
\centering
\includegraphics[width=0.9\textwidth]{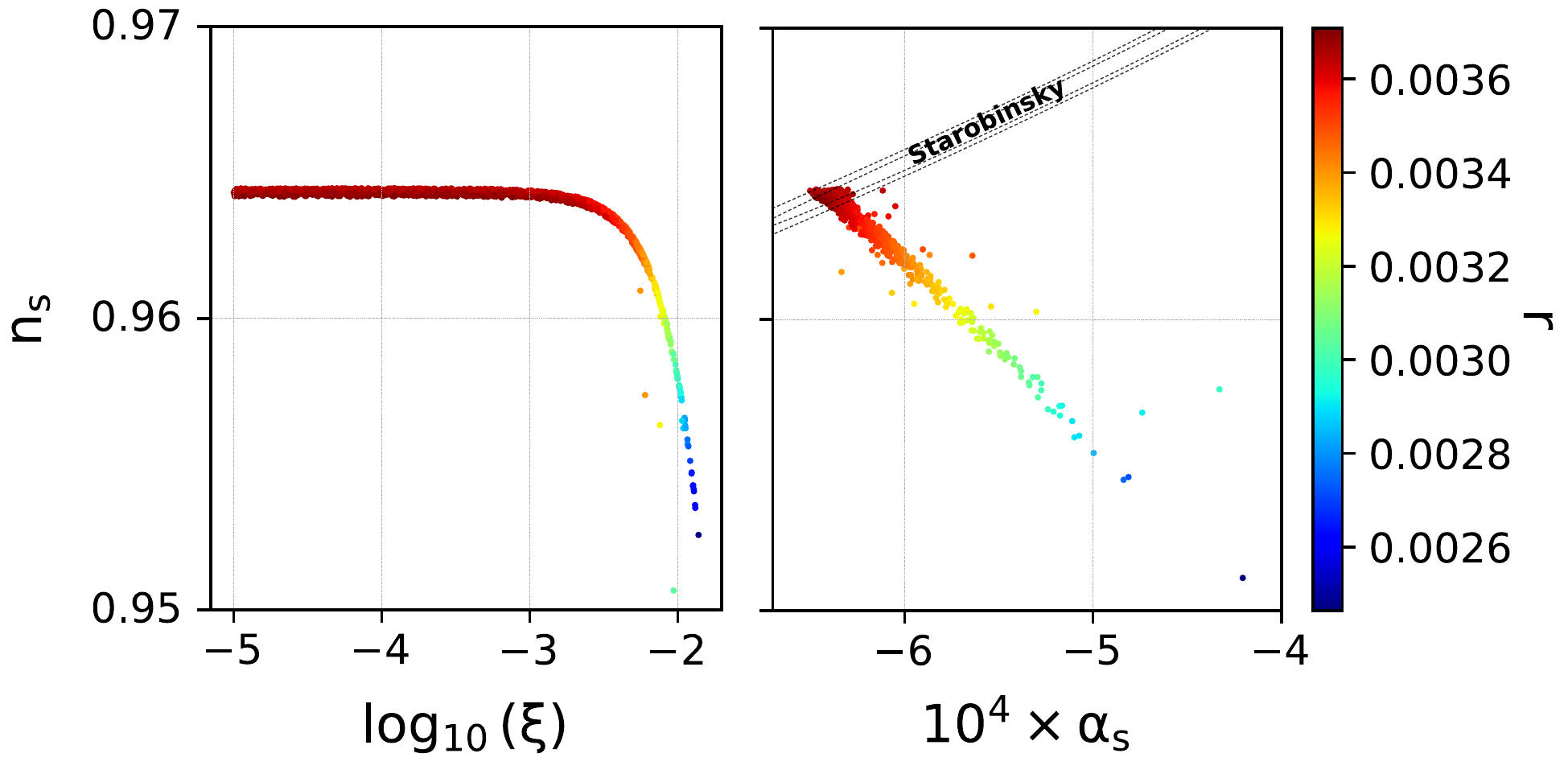}
\caption{\textbf{\textit{Left panel:}} 2D scatter plot in the $\log_{10}\xi$-$n_s$ plane [with $\xi$ controlling the strength of the $\phi^2R$ term in the Jordan frame action, see Eq.~(\ref{eq:actionjf})] colored by the value of the tensor-to-scalar ratio.\\ \textbf{\textit{Right panel:}} 2D scatter plot in the $10^4\alpha_s$-$n_s$ plane (with $\alpha_s=dn_s/d\log k$ the running of the spectral index) colored by the value of the tensor-to-scalar ratio, for the scale-invariant model studied in this work. The predictions in the $n_s$-$10^4\alpha_s$ plane are compared to those from Starobinsky inflation (contours given by the dotted curves, not colored by $r$).}
    \label{fig:scatter_plots}
\end{figure}

On the other hand, the parameter $\xi$ (controlling the strength of the non-minimal coupling $\phi^2R$) significantly influences both the scalar tilt $n_s$ and the amplitude of the primordial tensor power spectrum through the tensor-to-scalar ratio $r$. In the left panel of Fig.~\ref{fig:scatter_plots}, we observe that values of $\xi \sim 10^{-2}$ lead to a shift towards smaller $n_s \sim 0.95$, while for $\xi \lesssim 10^{-3}$ we converge to a flat plateau around $n_s \sim 0.965$, consistent with the \textit{Planck} results.~\footnote{There are two important caveats to this statement. In first place, CMB data from the Atacama Cosmology Telescope are in principle consistent with $n_s=1$, at the cost of some degree of tension with \textit{Planck}~\cite{ACT:2020frw,ACT:2020gnv,Handley:2020hdp,Giare:2022rvg,Calderon:2023obf}. Next, the quoted value of $n_s$ has been inferred within the $\Lambda$CDM model. However, models of early-time new physics invoked to address the Hubble tension~\cite{DiValentino:2021izs,Perivolaropoulos:2021jda,Schoneberg:2021qvd,Abdalla:2022yfr,Hu:2023jqc} typically lead to much higher inferred values of $n_s$, up to $n_s \approx 1$ (see e.g.\ Refs.~\cite{Poulin:2018cxd,Ye:2020btb,Ye:2021nej,Jiang:2021bab,Hazra:2022rdl,Jiang:2022uyg,Jiang:2022qlj,Poulin:2023lkg,Jiang:2023bsz,Peng:2023bik} for concrete examples). The reason is that such an increase in $n_s$ can absorb part of the shifts in the CMB spectrum related to an enhanced early integrated Sachs-Wolfe effect, as well as additional modifications in the damping tail~\cite{Poulin:2018cxd,Poulin:2023lkg,Lin:2019qug,Vagnozzi:2021gjh,Vagnozzi:2023nrq}. In light of these caveats, we caution against drawing too strong conclusions from the value of $n_s$ inferred by \textit{Planck} within $\Lambda$CDM~\cite{DiValentino:2018zjj}. See also the recent work of Ref.~\cite{Giare:2024akf} where this point was explored in great detail by one of us.} Therefore, to maintain consistency for what concerns the derived value of $n_s$, excessively high values of $\xi$ are not viable, as they would force us into a region of parameter space where $n_s$ becomes too small to remain in good agreement with \textit{Planck}. These considerations translate into the 95\%~C.L. upper limit $\xi<0.00142$. It is interesting to note that these constraints exclude $\xi=1$ at very high significance. Our choice of the $\phi^2R$ term in Eq.~(\ref{eq:actionjf}) having a coupling $\xi/6$ implies that $\xi=1$ is the conformal value for the coupling. Therefore, our results exclude conformal invariance within this specific model -- this is not necessarily surprising, given that conformal invariance is stronger than scale-invariance.\footnote{It is worth noting that the low-$n_s$ tail is more extended than the heavier high-$n_s$ tail. This leads to the somewhat peculiar shapes observed in the contours in Fig.~\ref{fig:triangularplot}. As discussed in the text and evident from the same Figure, lower values of $n_s$ are correlated with higher values of $\xi$, and the other way around. Therefore, the heavier high-$n_s$ tail reflects the lower edge in $\log_{10}\xi$ prior (which we set at $\log_{10}\xi > -5$). Of course, the choice of this edge is somewhat arbitrary since we cannot sample all the way down to $\xi=0$ once we choose to sample $\log_{10}\xi$. Nevertheless, as stressed above and in the text (see also the empirical convergence test discussed in footnote~10), our results are robust and converged, albeit not in the usual MCMC sense.} Moreover, we note that the allowed values of $\xi$ are much smaller than those required for the non-minimal coupling in Higgs inflation~\cite{Bezrukov:2007ep,Bezrukov:2010jz,Rubio:2018ogq}, and therefore our model does not suffer from the potential issues with unitarity violation raised in the context of Higgs inflation~\cite{Barbon:2009ya,Burgess:2010zq,Bezrukov:2010jz,Calmet:2013hia,Salvio:2015kka} (although we stress that a direct comparison between the two models is not fair, given that our scalar is not the Higgs).

When it comes to the parameter $\Omega$, which we recall controls the strength of the $\sinh^4(\rho)$ term in the potential within the $(\rho,\chi)$ field representation, or equivalently (once $\alpha$ and $\xi$ are given) the strength of the quartic term in the JF action, it is essential to clarify the choice of prior range within which we vary it: $\Omega \in [\xi^2, 2\xi^2]$. This prior is motivated by semi-analytical arguments which suggest that, for each given value of $\xi$, in order to prevent eternal inflation, $\Omega$ must be confined to a specific range of values $\Omega \in [\xi^2, 1.15\xi^2]$~\cite{Ghoshal:2022qxk}. Our analysis perfectly confirms this result numerically. In particular, we observe that when $\Omega$ deviates from the above interval, our numerical evolution leading to eternal inflation occurring. As a result, we adopt the prior reported in Tab.~\ref{tab:results} which, in order to be more conservative, allows more freedom compared to the tighter range obtained analytically and reported above. That being said, just as $\alpha$ and $\xi$, $\Omega$ too influences the inflationary potential and its related observables, as can be seen in Eq.~(\ref{eq:potentialrhochi}). This leads to strong correlations with the other parameters (both the model parameters and the inflationary observables) as observed in Fig.~\ref{fig:triangularplot}. In this case, we set a two-tailed constraint $\Omega = 0.93^{+0.72}_{-2.8} \times 10^{-5}$.

We conclude this subsection with some final remarks regarding the predictability of our scale-invariant model and the peculiarity of our methodology. The latter lies in our initial assumption of a specific model at the beginning of our analysis, differing significantly from the conventional approach in the literature. Typically, predictions of inflationary models are tested against cosmological data by superimposing theoretical curves (at fixed benchmark values for the fundamental parameters of the model) onto pre-obtained 2D marginalized probability contours in the $n_s$-$r$ plane. An illustrative example of this widespread approach can be appreciated in Fig.~8 of the well-known \textit{Planck} 2018 paper on inflation~\cite{Planck:2018jri}. In contrast, a distinct advantage of our method (adopted, among others, by some of us in Refs.~\cite{Giare:2023wzl,Giare:2024sdl}) is that we can derive precise model-dependent predictions for inflationary parameters, such as the tensor amplitude $r$ and the running of the spectral index $\alpha_s$. Once a specific inflationary model is assumed, the values of these quantities are mostly fixed as a result of consistency relations among different inflationary parameters.

The above considerations are evident from the correlations in the three-dimensional parameter space spanned by $n_s$, $r$, and $\alpha_s$ shown in the right panel of Fig.~\ref{fig:scatter_plots}. We observe that, within our model, more negative values of $\alpha_s$ are only possible when $n_s \sim 0.965$ and $r \sim 0.0036$. This region of parameter space corresponds to the regime where the effects of the non-minimal coupling $\phi^2R$ controlled by $\xi$ are negligible, and we therefore approach the regime of Starobinsky inflation where the $R^2$ term dominates. Conversely, lower values of $n_s \lesssim 0.96$ imply smaller values of $r$ and larger/less negative (thus smaller in absolute value) values of $\alpha_s$, deviating significantly from the predictions of Starobinsky inflation. Notice also that these correlations are obtained after marginalizing over all the free parameters of the model ($\xi$, $\alpha$, and $\Omega$, as well as initial conditions for the fields). Consequently, they offer a much more comprehensive overview of the interrelations between inflationary parameters within our specific scenario. This allows us to better highlight the predictive power of scale-invariant inflation. For instance, we anticipate a non-zero value for the amplitude of primordial gravitational waves $r > 0.00332$ and the spectral index running $\alpha_s < 1.2 \times 10^{-4}$ (both at 95\%~C.L.). We would like to stress that these results represent \textit{model-dependent} predictions which can be tested in light of future CMB experiments~\cite{CMB-S4:2016ple,SimonsObservatory:2018koc,SimonsObservatory:2019qwx,CMB-S4:2020lpa,LiteBIRD:2022cnt} to either validate or rule out the model. On the other hand, one might wonder whether (and to what extent) these predictions can help discriminate between this and other competing models. The next subsection is dedicated to discussing both aspects in detail.

\subsection{Comparison to Starobinsky inflation}
\label{subsec:discussion}

As we emphasized in the previous subsection, our observational constraints are derived by assuming the specific model of scale-invariant inflation in the analysis and testing its theoretical predictions against cosmological data, rather than merely superimposing theoretical curves onto previously obtained 2D marginalized $n_s$-$r$ contours. Consequently, each and every bound and correlation observed among inflationary parameters should be considered a prediction specific to our particular model. A natural question that arises is therefore whether and to what extent these predictions can be distinguished from other benchmark models, such as Starobinsky inflation~\cite{Starobinsky:1980te} or its $\alpha$-attractor extension~\cite{Kallosh:2013hoa,Kallosh:2013yoa,Galante:2014ifa,Kallosh:2015lwa,Kallosh:2022feu} (see also Refs.~\cite{Roest:2015qya,Linde:2015uga,Linder:2015qxa,Scalisi:2015qga,Odintsov:2016vzz,Braglia:2020bym,Rodrigues:2020fle,Shojaee:2020xyr,Bhattacharya:2022akq,Brissenden:2023yko}). In this regard, we note that Starobinsky inflation itself represents a scale-invariant model when the $R^2$ term dominates the inflationary dynamics (as is usually the case). One might therefore expect some degree of overlap in the predictions of the two models for certain ranges of parameters. We choose to perform the comparison of our model against Starobinsky inflation (and $\alpha$-attractor inflation) in light of the quasi-scale-invariant nature of the latter, in addition to its excellent agreement with current data, and its often being taken as a benchmark when studying the capabilities of future CMB experiments. However, we stress that a similar exercise can equally well be performed with other models.

\begin{figure}
\centering
\includegraphics[width=0.5\textwidth]{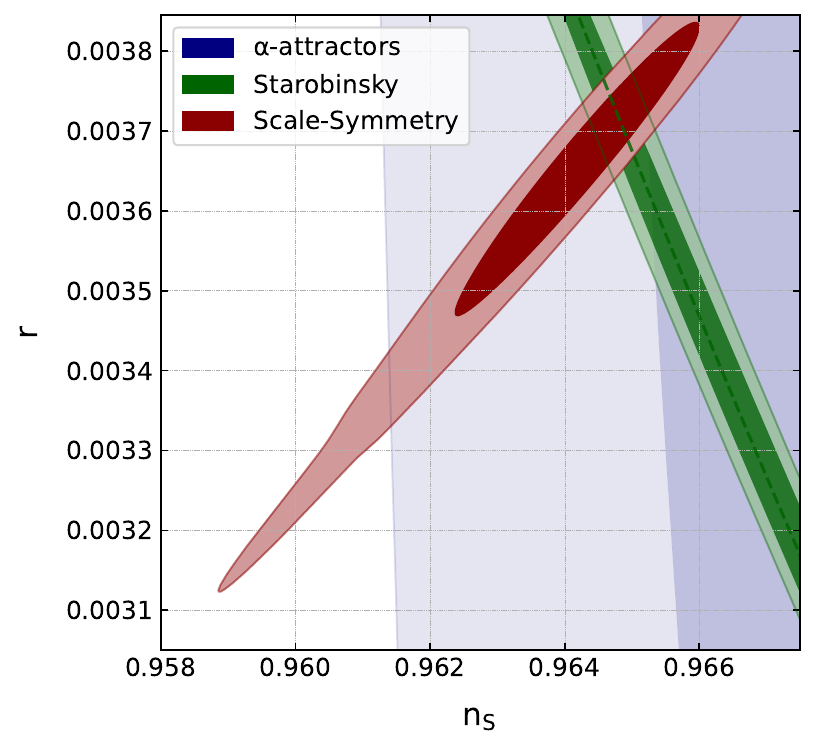}
\caption{2D contours in the $n_s$-$r$ plane for the scale-invariant model studied in this work (red contours), compared to those of Starobinsky inflation (green contours) and $\alpha$-attractors (light blue contours), obtained in light of observations from the \textit{Planck} 2018 legacy release and from the \textit{BICEP/Keck} collaboration (\textit{BICEP2}, \textit{Keck Array} and \textit{BICEP3} observations up to 2018) and sampling procedure described in Sec.~\ref{subsec:methods}.}
\label{fig:ns_r}
\end{figure}

To address these questions, in  Fig.~\ref{fig:ns_r} we compare marginalized contours in the $n_s$-$r$ plane for our scale-invariant inflationary model (red contours) against those obtained for Starobinsky inflation (green contours) and $\alpha$-attractor inflation (light blue contours). In the latter two cases, the predictions are derived following the methodology outlined in Ref.~\cite{Giare:2023wzl} where, broadly speaking, the universal predictions for the inflationary parameters:
\begin{eqnarray}
n_s\simeq 1 -\sqrt{\frac{r}{3\alpha}}\,,
\label{eq:alphaAtt}
\end{eqnarray}
with $\alpha=1$ corresponding to Starobinsky inflation, are assumed in the cosmological model, and constraints are derived from the same dataset analyzed for the scale-invariant inflationary model studied here, in order for the comparison to be meaningful. Further details on this methodology, as well as information on the priors assumed for the different parameters, can be found in Sec.~3 of Ref.~\cite{Giare:2023wzl}.

With reference to Fig.~\ref{fig:ns_r}, although the predictions for the three models overlap within a fairly large region of parameter space, a number of interesting differences can be noted. First and foremost, the red contours indicate a positive correlation between the amplitude of tensor modes and the scalar spectral index in the scale-invariant model of inflation. In other words, higher values of $r$ are correlated with higher values of $n_s$. This is not the case in Starobinsky inflation, where these two parameters are related by Eq.~(\ref{eq:alphaAtt}) with $\alpha=1$, thereby leading to a behavior which is exactly the opposite. This relation clearly imposes a negative correlation, as higher values of $r$ lead to lower values of $n_s$. Due to the different correlation between $n_s$ and $r$ in the two models, the respective marginalized probability contours are rotated relative to each other, and a substantial portion of the parameter space falling within the 68\%~C.L. region in scale-invariant inflation would be excluded in Starobinsky inflation at a statistical level (largely) exceeding 95\%~C.L. (and viceversa). Considering the more general $\alpha$-attractor predictions -- i.e.\ Eq.~(\ref{eq:alphaAtt}) with $\alpha$ free to vary -- the correlation between $n_s$ and $r$ is basically lost due to marginalization upon the additional parameter $\alpha$. Given that when $n_s$ is fixed, $\alpha$ only affects the amplitude of tensor modes (interpolating vertically along the $n_s$-$r$ plane between the predictions of Starobinsky inflation and those of the monomial $\propto \phi^2$ model), we are left with a substantial freedom to accommodate different values of $r$ by appropriately changing $\alpha$, which reflects in very broad contours once $\alpha$ is marginalized over.

Summing up, the increased freedom in $\alpha$-attractor inflation complicates the discrimination between the models, given the much broader 2D probability contours obtained in this case. In contrast, the differences between scale-invariant inflation and Starobinsky inflation appear substantial enough to raise questions about \textit{i)} the physical nature of the differences, and \textit{ii)} conclusions one can draw about these models in light of forthcoming, more precise CMB observations, both satellite- and ground-based. Our observations are as follows:
\begin{enumerate}
\item Regarding the first point, in the previous subsection we argued that $\xi$ is the parameter which most prominently affects constraints in the $n_s$-$r$ plane. Taking another look at the left-side panel of Fig.~\ref{fig:scatter_plots}, we highlight once more how the predictions of the Starobinsky model are, in fact, fully recovered in the limit $\xi \to 0$, when the dynamics are driven by the $R^2$ term. In the regime of negligible $\xi$, we converge to the plateau described by $n_s\approx 0.965$ and $r\approx 0.037$, which corresponds precisely to the region where the red and green contours in Fig.~\ref{fig:ns_r} overlap. In contrast, as the value of $\xi$ becomes sufficiently high, the shift towards lower values of the spectral index and tensor amplitude forces us into a region of the $n_s$-$r$ plane inaccessible to Starobinsky inflation. The same pattern is also evident from the right panel of Fig.~\ref{fig:scatter_plots}, wherein we see that as long as the contribution of $\xi$ remains negligible (left-most side of the panel), the predictions of scale-invariant inflation (colored points) match those of Starobinsky inflation (dashed contours) also in the 3D plane spanned by $n_s$, $r$, and $\alpha_s$.

The fact that the predictions of scale-invariant inflation reduce to Starobinsky in the limit $\xi\to0$ can also be proven through semi-analytical arguments, providing direct evidence for the robustness of our numerical results. Indeed, let us consider the potential slow-roll parameters~\cite{Martin:2013tda}, solely a function of the field $\rho$:
\begin{equation}
\epsilon(\rho) = \dfrac{M_p^2}{2}\left(\dfrac{V_{,\rho}}{V}\right)^2\,, \quad \quad \eta(\rho) = M_p^2 \dfrac{V_{,\rho \rho}}{V}\,,
\label{eq:epsilonrhoetarho}
\end{equation}
from which we obtain the spectral indices as:
\begin{equation}
n_s(\rho) \simeq 1-6\epsilon(\rho) + 2\eta(\rho)\,, \qquad r \simeq 16 \epsilon(\rho)\,.
\label{eq:nsrhoetarho}
\end{equation}
These expressions are easily computed analytically and greatly simplified within the limit $\Omega\to\xi^2$, which is motivated by virtue of the above discussion. By further taking the limit $\xi\to 0$, we obtain:
\begin{eqnarray}
n_s &\simeq& 1-\dfrac{8}{3}\xi \cosh \left ( \sqrt{\dfrac{2}{3}}\dfrac{\rho}{M_p} \right ) \,, \\
r &\simeq& \dfrac{64}{3}\xi^2 \sinh^2 \left ( \sqrt{\dfrac{2}{3}}\dfrac{\rho}{M_p} \right ) \,.
\label{eq:nsrlimitnoxi}
\end{eqnarray}
The two above equations can be combined to obtain: 
\begin{equation}
n_s \simeq 1-\dfrac{1}{3}\sqrt{3r + 64\xi^2}\simeq 1-\sqrt{\dfrac{r}{3}}\,,
\label{eq:nsrstarobinsky}
\end{equation}
which is precisely the same relation found in Starobinsky inflation, see e.g.\ Eq.~(\ref{eq:alphaAtt}) with $\alpha=1$. However, as we stressed earlier, as soon as $\xi \neq 0$ and therefore the non-minimal coupling $\phi^2R$ is turned on, the predictions of the two models differ, as is clear from Fig.~\ref{fig:ns_r}.

\item Concerning future observations, we anticipate that the upcoming generation of CMB experiments, including the ground-based Simons Observatory (SO)~\cite{SimonsObservatory:2018koc,SimonsObservatory:2019qwx} and CMB-S4~\cite{CMB-S4:2016ple,CMB-S4:2020lpa}, and the LiteBIRD satellite~\cite{LiteBIRD:2022cnt}, will significantly improve our ability to constrain the tensor amplitude, hopefully leading to a detection of primordial tensor modes. In some cases, these surveys are expected to reduce the observational uncertainties in the value of the spectral index as well. Just to quote a few concrete values, assuming the usual power-law parametrization for primordial spectra, the official collaboration forecasts suggest that SO should achieve a sensitivity of $\sigma(r)\sim 0.003$ on tensor modes, along with an improvement in the constraining power on the scalar spectral index of up to $\sigma(n_s)\sim 0.003$. On the other hand, CMB-S4 is expected to reach $\sigma(r)\sim 0.001 - 0.007$ (with marginal dependency on foreground modelling, as discussed in Ref.~\cite{CMB-S4:2020lpa}) and $\sigma(n_s)\sim 0.002$. Additionally, LiteBIRD is also anticipated to have a similar sensitivity of $\sigma(r)\lesssim 0.001$, although its impact on the scalar spectral index is expected to be limited (this is not surprising, given that LiteBIRD is targeting the large angular scale CMB polarization signal).

The sensitivity forecasts anticipated from upcoming experiments quoted above could provide crucial insights into scale-invariant inflation. First and foremost, optimistically assuming that the model provides an accurate description of the inflationary Universe, based on the results obtained from current data, we \textit{predict} the presence of a primordial gravitational wave background with an amplitude $r \gtrsim 0.003$. These values of the tensor amplitude are sufficiently large to be visible from all the above-mentioned experiments. For instance, CMB-S4 is expected to detect primordial gravitational waves with $r>0.003$ at a statistical significance of up to $5\sigma$ or larger. In case of lack of detection, the same experiment should instead set an upper limit $r<0.001$ at a 95\%~C.L.: it is therefore clear that, extrapolating the constraints from current data, failure to detect $r$ (and therefore primordial tensor modes) by CMB-S4 would strongly contradict the predictions of scale-invariant inflation, essentially ruling out the model. Similar considerations can be drawn for SO and LiteBIRD.

That being said, in the optimistic scenario where future experiments will provide a detection of $r$, the question of whether a combined inference of $n_s$ and $r$ will be sufficient to discriminate between competing models remains open. On the one hand, we expect these experiments to significantly narrow down the region in the $n_s$-$r$ plane where all inflationary models should lie compared to what we can currently infer from the analysis of \textit{Planck} and \textit{BICEP/Keck} data. Therefore, given the differences surrounding the predictions of the different models observed in Fig.~\ref{fig:ns_r}, one might be tempted to conclude that future data can eventually help discriminate between Starobinsky and scale-invariant inflation. While this is in part true, it is important to note that predictions in this plane crucially depend on the value of $\xi$. Assuming that large values of $\xi$ are realized in Nature, we note that the differences in $n_s$ between these two models can be at most as large as $\Delta n_s \sim 0.006$ when $r \sim 0.032$. Although these differences could be significant when compared to the sensitivities to $n_s$ mentioned above, it should be noted that in the more realistic case where $\xi$ approaches small values, the differences become much smaller for higher values of the amplitude of tensor modes. In general, even within a very optimistic setup, these differences never exceed $2-3$ times the forecasted observational uncertainties. In light of these considerations, it appears quite unlikely that future data can provide conclusive evidence in favor of one model over the other at high statistical significance. The situation could change if, as discussed in footnote~7, an eventual widely agreed upon solution to the Hubble tension ends up pushing us towards a region in $n_s$ parameter space far from the one currently favored within $\Lambda$CDM: in this case, however, a significant rethinking of the inflationary paradigm might be required (see e.g.\ Refs.~\cite{Vallinotto:2003vf,Starobinsky:2005ab,Takahashi:2021bti,DAmico:2021fhz,Lin:2022gbl,Ye:2022efx,Fu:2023tfo}). At any rate, we defer a more complete forecast for the power of upcoming CMB experiments to discriminate between scale-invariant inflation and competing models such as Starobinsky inflation to future work.
\end{enumerate}

\section{Conclusions}
\label{sec:conclusions}

There are strong top-down and bottom-up motivations underlying the idea of scale-invariance as a fundamental symmetry of Nature. On the theoretical side, quantum scale symmetry is a highly predictive guiding principle, beyond renormalizability, for a quantum theory of gravity. On the phenomenological side, from both the particle physics and cosmological perspectives, equally strong motivation for classical scale-invariance exists. In particular, (quasi-)scale-invariant inflationary models (such as Starobinsky inflation) naturally possess features which allow them to accommodate the observed small amount of anisotropies and stringent limits on the amplitude of primordial tensor modes. Driven by these considerations, in the present work we have studied in detail a scale-invariant model of inflation presented earlier by one of us~\cite{Rinaldi:2015uvu}: within the model, whose action features terms quadratic in curvature and a scalar field non-minimally coupled to gravity, and obviously contains no explicit mass scale (see Eq.~(\ref{eq:actionjf}) in the Jordan frame), inflation takes place during the transition between two de Sitter regimes, when spontaneous breaking of scale-invariance occurs and a mass scale ultimately identified with the Planck mass emerges (see also Refs.~\cite{Salvio:2014soa,Salvio:2017qkx}). It is important to stress that the action given in Eq.~(\ref{eq:actionjf}) is meant to be considered as exact, rather than as an effective expansion of a yet unknown theory. In other words, scale-invariance here is implemented by a non-perturbative theory -- this is the main difference with respect to Starobinsky inflation, where scale-invariance is restored only at large curvature values.

In this work, we go significantly beyond earlier semi-analytical works, which studied the model's predictions for the inflationary parameters $n_s$ and $r$~\cite{Tambalo:2016eqr,Ghoshal:2022qxk,Rinaldi:2023mdf}. Our model predictions are derived by solving the full numerical dynamics of the system, making use of a method recently developed by some of us to study generic multi-field inflationary models with potentially non-trivial field-space metric~\cite{Giare:2023kiv}, and are confronted against the latest \textit{Planck} and \textit{BICEP/Keck} CMB observations to obtain the first robust constraints on the free parameters of the model. These are particularly robust given their having been derived assuming the specific model of scale-invariant inflation in the analysis, rather than merely superimposing theoretical curves onto pre-obtained 2D marginalized contours in the $n_s$-$r$ plane (and, by extension, imposing theory-driven rather than observation-driven priors).

Our main findings can be summarized as follows:
\begin{enumerate}
\item we find a very tight upper limit on the parameter $\xi<0.00142$ (at 95\%~C.L.) controlling the strength of the non-minimal coupling $\phi^2R$ in the Jordan frame action [see Eq.~(\ref{eq:actionjf})], which excludes the conformal value $\xi=1$ at extremely high significance;
\item for the other two model parameters, $\alpha$ and $\Omega$ related respectively to the quadratic in curvature ($R^2$) and quartic ($\phi^4$) terms in the Jordan frame action [see Eqs.~(\ref{eq:actionjf},\ref{eq:omega})], we infer $\alpha \sim 2 \times 10^{10}$ and $\Omega \sim 10^{-5}$ -- in particular, our constraints on $\xi$, $\alpha$, and $\Omega$ corroborate the results obtained in earlier semi-analytical works;
\item we numerically confirm that, despite the model being on paper a two-field one, its dynamics are truly driven by only one field, as is clear from the discussion in Sec.~\ref{subsec:redefinition}, and as a direct consequence of scale-invariance;
\item we prove that the sensitivity to the initial conditions for the two fields is extremely limited -- when explicitly sampling over these initial conditions, we have in fact found them to be unconstrained;
\item we analytically and numerically demonstrate that the model predicts vanishing entropy perturbations, again a direct consequence of scale-invariance;
\item we semi-analytically compute the predicted level of non-Gaussianity in the squeezed limit, finding it to be very small [${\cal O}(0.1)$ or smaller] and consistent with CMB observations -- this confirms once more that the model dynamics are effectively single-field;
\item we illustrate how the model's \textit{predictions} (we use this term in light of our numerical approach of assuming the specific model when confronting it against observations, see above) quantitatively differ from those of Starobinsky inflation, for what concerns the directions of the mutual correlations between $n_s$, $r$, and the running of the spectral index $\alpha_s$, explaining by means of semi-analytical arguments the origin of these differences;
\item similarly, we \textit{predict} $r \gtrsim 0.003$, which is well within the expected sensitivity of upcoming CMB experiments -- conversely, lack of detection of primordial tensor modes at the level we predict would rule out our scale-invariant inflationary model;
\item finally, we have qualitatively argued that despite the aforementioned differences with respect to Starobinsky inflation, it will be hard for upcoming CMB experiments to discriminate between the two models at a high significance level.
\end{enumerate}
The above findings corroborate previous semi-analytical studies, thereby placing the model and its earlier studies on a more robust footing.

There are several potentially interesting avenues for follow-up work. On the theoretical side, the key aspect of the model which remains to be studied is the potential production of primordial black holes, together with predictions for the associated stochastic background of gravitational waves, particularly in light of the signal recently detected by pulsar timing array experiments. A more detailed study of the (p)reheating dynamics is also in order, as a follow-up of what was done earlier in Ref.~\cite{Rinaldi:2015uvu}. Moreover, while we have focused on a classically scale-invariant action, quantum corrections can spoil scale-invariance, on the one hand possibly leading to entropy perturbations and therefore an additional avenue of observational constraints, and on the other hand potentially leading to features which may facilitate primordial black hole production: these aspects require a dedicated study which goes well beyond the scope of the present work (and would presumably require including the square of the Weyl tensor in the action). In addition, when endowed with non-zero spatial curvature, our model can naturally support a bouncing scenario whose dynamics are yet to be studied in detail. On the observational side, it would be desirable to go beyond our semi-qualitative forecast. For instance, a fully-fledged forecast would entail the generation of mock CMB data assuming our specific model, with the instrumental specifications for future experiments entering instead in the (mock) likelihood. Moreover, a more detailed study of the primordial three-point correlation function (in particular considering other configurations beyond the squeezed one) would be very valuable, and would allow for a direct comparison to bispectrum measurements in the CMB and in galaxy surveys. We leave a detailed exploration of these and other interesting points to follow-up work.

To sum up, in the present work we have performed the first robust comparison of scale-invariant inflation against current precision cosmological observations from the CMB, at least for what concerns the use of multi-field dynamics to numerically follow the evolution of the two fields in scale-invariant scenarios. Our findings confirm that the model is in extremely good health, and indicate that upcoming CMB observations (such as SO, CMB-S4, or LiteBIRD) will rule it out if values of the tensor-to-scalar ratio $r \gtrsim 0.003$ are excluded -- in this sense, considering also its strong theoretical motivation, we feel that this model provides another interesting benchmark for tests of inflation from future CMB experiments. Our work further reinforces the potential key role of scale-invariance as being the symmetry underlying the inflationary paradigm, while also being an important theoretical guiding principle.

\section*{Acknowledgments}
M.D.A.\ and W.G.\ are grateful to Carsten van de Bruck and Eleonora Di Valentino for valuable suggestions. C.C.\ acknowledges the hospitality of CERN, where the final part of this work was carried out. W.G.\ is supported by the Lancaster–Sheffield Consortium for Fundamental Physics under STFC grant ST/X000621/1. S.V.\ acknowledges support from the University of Trento and the Provincia Autonoma di Trento (PAT, Autonomous Province of Trento) through the UniTrento Internal Call for Research 2023 grant ``Searching for Dark Energy off the beaten track'' (DARKTRACK, grant agreement no.\ E63C22000500003). C.C., M.R., and S.V.\ acknowledge support from the Istituto Nazionale di Fisica Nucleare (INFN) through the Commissione Scientifica Nazionale 4 (CSN4) Iniziativa Specifica ``Quantum Fields in Gravity, Cosmology and Black Holes'' (FLAG). This publication is based upon work from the COST Action CA21136 ``Addressing observational tensions in cosmology with systematics and fundamental physics'' (CosmoVerse), supported by COST (European Cooperation in Science and Technology). M.D.A.\ and W.G.\ acknowledge IT Services at the University of Sheffield for the provision of services for High Performance Computing.

\bibliographystyle{JHEP}
\bibliography{Scaleinvariant}

\end{document}